\documentclass[11pt,final]{article}
\usepackage{fullpage}
\usepackage[numbers,compress]{natbib}

\usepackage{microtype}
\usepackage{graphicx}
\usepackage{subfigure}
\usepackage{booktabs} 
\usepackage{forest}

\usepackage{amsfonts}       
\usepackage{nicefrac}       
\usepackage{xspace}

\usepackage{algorithm}
\usepackage[noend]{algorithmic}

\usepackage{enumitem}
\usepackage{xcolor}
\usepackage{gensymb} 
\usepackage{graphicx}
\usepackage{amsmath}
\usepackage{dsfont}
\usepackage{multirow} 
\usepackage{amssymb} 
\usepackage{amsthm}
\usepackage{mathtools}

\makeatletter
\let\original@algocf@latexcaption\algocf@latexcaption
\long\def\algocf@latexcaption#1[#2]{%
  \@ifundefined{NR@gettitle}{%
    \def\@currentlabelname{#2}%
  }{%
    \NR@gettitle{#2}%
  }%
  \original@algocf@latexcaption{#1}[{#2}]%
}
\makeatother

\usepackage{eqparbox}

\newcommand{\mainicalg}{\hyperref[mainicalg]{IC-FEE}\xspace}
\newcommand{\mainalg}{\hyperref[mainalg]{FEE}\xspace}
\newcommand{\algname}{\hyperref[mainalg]{FEE}\xspace}
\newcommand{\mainicepalg}{\hyperref[mainicepalg]{IC-EP-FEE}\xspace}
\newcommand{\gre}{\textsc{Greedy}\xspace}
\newcommand{\fulexp}{\textsc{full-exploration}\xspace}

\newcommand{\lIF}[2]{\textbf{if} #1 \textbf{then} #2}
\newcommand{\lELSE}[1]{\textbf{else}, #1}

\newenvironment{proofof}[1]{\begin{proof}[\textnormal{\textbf{Proof of #1}}]}{\end{proof}} 

\newcount\Comments  
\newcount\Includeappendix 
\newcount\Putacknowledgement 

\definecolor{darkgreen}{rgb}{0,0.6,0}
\newcommand{\kibitz}[2]{\ifnum\Comments=1{\color{#1}{#2}}\fi}

\newcommand{\omer}[1]{\kibitz{blue}{[Omer :#1]}}

\newcommand{\red}[1]{\color{red}{#1}}

\makeatletter
\newcommand{\RemoveAlgoNumber}{\renewcommand{\fnum@algocf}{\AlCapSty{\AlCapFnt\algorithmcfname}}}
\newcommand{\RevertAlgoNumber}{\algocf@resetfnum}
\makeatother

\newcommand{\mm}[1]{M^{(#1)}}
\newcommand{\tih}{\tilde h}

\newcommand{\eal}{{E^r_{rec}}}
\newcommand{\eopen}{{E^r_{open}}}
\newcommand{\eopt}{{E^r_{opt}}}

\newcommand{\OPT}{\textnormal{OPT}}
\newcommand{\OPTEA}{\textnormal{OPT}_{\textnormal{EAIR}}}
\newcommand{\OPTEP}{\textnormal{OPT}_{\textnormal{EPIR}}}
\newcommand{\OPTDEL}{\textnormal{OPT}_{\textnormal{DEL}}}

\newcommand{\mP}{\mathcal{P}}

\newcommand{\mA}{\mathcal{A}}

\newcommand{\mR}{\mathcal{R}}

\newcommand{\mS}{\mathcal{S}}

\newcommand{\ind}{\mathds{1}}

\newcommand{\R}{\mathbb{R}}
\newcommand{\N}{\mathbb{N}}
\newcommand{\defeq}{\stackrel{\text{def}}{=}}

\newtheorem{assumption}{Assumption}
\newtheorem{claim}{Claim}
\newtheorem{definition}{Definition}
\newtheorem{theorem}{Theorem}
\newtheorem{proposition}{Proposition}
\newtheorem{lemma}{Lemma}

\newtheorem{observation}{Observation}

\newcommand\bl[1]{\boldsymbol{ #1 } }

\newcommand\abs[1]{\left| #1  \right|}
\DeclareMathOperator*{\argmax}{arg\,max} 

\DeclareMathOperator{\E}{\mathbb{E}}

\DeclarePairedDelimiter{\ceil}{\lceil}{\rceil}

\newcommand{\mep}{M_{\text{EPIR}}}

\usepackage{hyperref}       
\hypersetup{
     colorlinks   = true,
	 citecolor = teal,
}

\title{Fiduciary Bandits}

\author{
\and Gal Bahar%
\thanks{%
    {Technion - Israel Institute of Technology (\url{bahar@campus.technion.ac.il})}}
\and Omer Ben{-}Porat%
\thanks{%
    {Technion - Israel Institute of Technology (\url{omerbp@campus.technion.ac.il})}, corresponding author}
\and Kevin Leyton{-}Brown%
\thanks{%
    {University of British Columbia, Canada (\url{kevinlb@cs.ubc.ca})}}
\and Moshe Tennenholtz%
\thanks{%
    {Technion - Israel Institute of Technology (\url{moshet@ie.technion.ac.il})}}
}

\begin{document}

\maketitle

\begin{abstract}
Recommendation systems often face exploration-exploitation tradeoffs: the system can only learn about the desirability of new options by recommending them to some user. Such systems can thus be modeled as multi-armed bandit settings; however, users are self-interested and cannot be made to follow recommendations. We ask whether exploration can nevertheless be performed in a way that scrupulously respects agents' interests---i.e., by a system that acts as a \emph{fiduciary}. 
More formally, we introduce a  model  in  which  a recommendation system faces an exploration-exploitation tradeoff under the constraint that it can never recommend any action that it knows  yields lower reward in expectation than an agent would achieve if it acted alone. Our main contribution is a positive result: an asymptotically optimal, incentive compatible, and \emph{ex-ante} individually rational recommendation algorithm.
\end{abstract}

\section{Introduction}\label{sec:intro}
Multi-armed bandits (henceforth MABs) \citep{Bubeck2012,CesaBianchi} is a well-studied problem domain in online learning. In that setting, several arms (i.e., actions) are available to a planner; each arm is associated with an unknown reward distribution, from which rewards are sampled independently each time the arm is pulled. The planner selects arms sequentially, aiming to maximize her sum of rewards. This often involves a tradeoff between \textit{exploiting} arms that have been observed to yield good rewards and \textit{exploring} arms that could yield even higher rewards. Many variations of this model exist, including stochastic \citep{abbasi2011improved,karnin2013almost}, Bayesian \citep{agrawal2012analysis,chapelle2011empirical}, contextual \citep{chu2011contextual,slivkins2014contextual}, adversarial \citep{auer1995gambling} and non-stationary  \citep{besbes2014stochastic,Levine2017} bandits.

This paper considers a setting motivated by recommender systems. Such systems suggest actions to agents based on a set of current beliefs and assess agents' experiences to update these beliefs. For instance, in navigation applications (e.g., Waze; Google maps) the system recommends routes to drivers based on beliefs about current traffic congestion. The planner's objective is to minimize users' average travel time. The system cannot be sure of the congestion on a road segment that no agents have recently traversed; thus, navigation systems offer the best known route most of the time and explore occasionally. Of course, users are not eager to perform such exploration; they are self-interested in the sense that they care more about minimizing their own travel times than they do about conducting surveillance about traffic conditions for the system.

A recent line of work \citep{Kremer2014,Mansour2015}, inspired by the viewpoint of algorithmic mechanism design \citep{nisan1999algorithmic, nisan2007algorithmic}, deals with that challenge by \textit{incentivizing exploration}---that is, setting up the system in such a way that no user would ever rationally choose to decline an action that was recommended to him. The key reason that it is possible to achieve this property while still performing a sufficient amount of exploration is that the planner has more information than the agents. At each point in time, each agent holds beliefs about the arms' reward distributions; the planner has the same information, but also knows about all of the arms previously pulled and the rewards obtained in each case. More specifically, \citet{Kremer2014} consider a restricted setting and devise an MAB algorithm that is \textit{incentive compatible} (IC), meaning that
whenever the algorithm recommends arm $i$ to an agent, the best response of the agent is to select arm $i$.

Although this approach explicitly reasons about agents' incentives, it does not treat agents \emph{fairly}: agents who are asked to explore receive lower expected rewards. More precisely, in their attempt to reach optimality (in the static setting) or minimize regret (in the online setting), these IC MAB algorithms are intentionally providing (\textit{a priori}) sub-optimal recommendations to some of the agents. In particular, some of the agents could be better off by not using the system and follow their \textit{default arm}--- the \textit{a priori} superior arm, which would be every agent's rational choice in the absence both of knowledge of other agents' experiences and of a trusted recommendation. Thus, it would be natural for agents to see the recommendations of such IC MAB algorithms as a betrayal of trust; they might ask  ``why should I trust a recommender that occasionally gives out recommendations it has every reason to believe could make me worse off?''

In this work, we explicitly suggest that a social welfare maximization standpoint might raise societal issues, harming the trust agents put in recommender systems. The central premise of this paper is that explore-and-exploit AI systems should satisfy \textit{individual guarantees}---guarantees that the system should fulfill for each agent \textit{independently from the other agents and their recommendations}. At the one end of the spectrum are current MAB algorithms---successful in maximizing welfare, but do not offer the slightest individual guarantee. At the other end is the \textit{fiduciary duty}: borrowed from law applications, it requires that the mechanism acts in the interest of its clients with all its knowledge. This is the strictest, and strongest, individual guarantee the system could provide. However, if we applied this standard, we would be left only with the mechanism that greedily picks the apparently best arm in each iteration. In some settings, perhaps this is the best that can be achieved; however, note that this mechanism is rarely able to learn anything. It is therefore natural to ask for an approach that enjoys both worlds---maximizing welfare while satisfying individual guarantees.

\paragraph{Our contribution} 
We explore a novel compromise between these two extreme points, which we call \textit{ex-ante} individual rationality (EAIR). To motivate it, we consider the benchmark reward of each agent to be that of the default arm: the reward agents would get if the recommender system is unavailable. A mechanism is EAIR if the reward of every recommendation it makes beats that benchmark in expectation, per the mechanism's knowledge. More technically, a mechanism is EAIR if any probability distribution over arms that it selects has expected reward that is always at least as great as the reward of the default arm, both calculated based on the mechanism's knowledge (which is more extensive than that of agents). While it is possible for the mechanism to sample a recommendation from a distribution that is \emph{a priori} inferior to the (realization of the) default arm, the agent receiving the recommendation is nevertheless guaranteed to realize expected reward weakly greater than that offered by the default arm. Satisfying this requirement makes a MAB algorithm more appealing to agents; we foresee that in some domains, such a requirement might be imposed as fairness constraints by authorities. 

Algorithmically, we focus on constructing optimal EAIR mechanisms. Our model is a bandit model with $K\geq 2$ arms and $n$ agents (rounds). Similarly to \citet{Kremer2014}, we assume that rewards are fixed but initially unknown. 

We consider two agent schemes. In the first part of the paper, we assume that agents follow recommendations, as in the classical MAB literature. This is the case if, e.g., agents are oblivious to some of the actions' desirability, unaware of the entire set of alternatives, or if the cognitive overload of computing expectations is high. The main technical contribution of this paper is an EAIR mechanism, which obtains the highest possible social welfare by any EAIR mechanism up to an additive factor of $o(\frac{1}{n})$. Due to our static setting (rewards are realized only once), following the wrong exploration policy for even one agent has detrimental effect on social welfare. The optimality of our mechanism, which we term Fiduciary Explore \& Exploit (\mainalg) and outline as Algorithm \ref{mainalg}, follows from a careful construction of the exploration phase. Our analysis uses an intrinsic property of the setting, which is further elaborated in Theorem \ref{thm:optimal policy}. 

Later on, in Section \ref{sec:ic}, we adopt a different agent scheme, which is fully aligned with the incentivizing exploration literature. We assume that agents are strategic and have (the same) Bayesian prior over the rewards of the arms. In this context, a mechanism is \textit{incentive compatible} (IC) if each agent's expected reward is maximized by the recommended action. We provide a positive result in this challenging case as well. Our second technical contribution is Incentive Compatible Fiduciary Explore \& Exploit (\mainicalg), which uses \mainalg as a black box, and is IC, EAIR and asymptotically optimal.
 
To complement this analysis, we also propose the more demanding concept of \textit{ex-post} individual rationality (EPIR). The EPIR condition requires that a recommended arm must never be \textit{a priori} inferior to the default arm given the planner's knowledge. The EAIR and EPIR requirements differ in the guarantees that they provide to agents and correspondingly allow the system different degrees of freedom in performing exploration. We design an asymptotically optimal IC and EPIR mechanism. Finally, we analyze the social welfare cost of adopting either EAIR or EPIR mechanisms.
\paragraph{Related work}

Background on MABs can be found in \citet{CesaBianchi} and a recent survey \cite{Bubeck2012}. Despite that many works address MAB rounds as interacting agents, \citet{Kremer2014} is the first work of which we are aware that suggests that vanilla algorithms should be modified to deal with agents due to human nature and incentives. The authors considered two deterministic arms, a prior known both to the agents and the planner, and an arrival order that is common knowledge among all agents, and presented an optimal IC mechanism. \citet{LeeCohen2019} extended this optimality result to several arms under further assumptions. This setting has also been extended to regret minimization \citep{Mansour2015}, social networks \citep{Bahar2016,bahar2019recommendation}, and heterogeneous agents \citep{chen18a,immorlica2019bayesian}. All of this literature disallows paying agents; monetary incentives for exploration are discussed in e.g., \citep{chen18a,Frazier2014Kleinberg}. None of this work considers the orthogonal, societal consideration of individual rationality constraint as we do here.

Our work also contributes to the growing body of work on fairness in Machine Learning \citep{BenPoratT18,dwork2012fairness,hardt2016equality,liu2018delayed}. In the context of MABs, some recent work focuses on fairness in the sense of treating \textit{arms} fairly. In particular, \citet{liu2017calibrated} aim at treating similar arms similarly and  \citet{MatthewKearnsMorgensternRothNIPS2016} demand that a worse arm is never favored over a better one despite a learning algorithm's uncertainty over the true payoffs. 
Finally, we note that the EAIR requirement we impose---that agents be guaranteed an expected reward at least as high as that offered by a default arm---is also related to the burgeoning field of safe reinforcement learning \citep{garcia2015comprehensive}.
\section{Model}\label{sec:model}
Let $A=\{a_1,\dots a_K\}$ be a set of $K$ arms (actions). Rewards are deterministic but initially unknown: the reward of arm $a_i$ is a random variable $X_i$, and $(X_i)_{i=1}^K$ are mutually independent. We denote by $R_i$ the observed value of $X_i$. To clarify, rewards are realized only once; hence, once $R_i$ is observed, $X_i=R_i$ for the rest of the execution. Further, we denote by $\mu_i$ the expected value of $X_i$, and assume for notational convenience that $\mu_1 \geq \mu_2 \geq\dots \geq \mu_K$. We also make the simplifying assumption that the rewards $(X_i)_{i=1}^K$ are fully supported on the set $[H]^+ \defeq\{0,1,\dots ,H\}$, and refer to the continuous case in Section \ref{discussions}.  

There are $n$ agents, who arrive sequentially. We denote by $a^l$ the action of the agent arriving at stage $l$. The reward of the agent arriving at stage $l$ is denoted by $R^l$, and is a function of the arm she chooses. For instance, by selecting arm $a_r$ the agent obtains $R^l(a_r)=X_r$. Agents are fully aware of the distribution of $(X_i)_{i=1}^K$. Each and every agent cares about her own reward, which she wants to maximize. 

A mechanism is a recommendation engine that interacts with agents. The input for the mechanism at stage $l$ is the sequence of arms pulled  and rewards received by the previous $l-1$ agents. The output of the mechanism is a recommended arm for agent $l$. Formally, a mechanism is a function $ M:\bigcup_{l=1}^n \left(A\times \R_+ \right)^{l-1} \rightarrow \Delta(A); $
of course, we can also define a deterministic notion that maps simply to $A$. The mechanism has a global objective, which is to maximize agents' social welfare $\sum_{l=1}^n R^l(a^l)$. 

We consider two agent schemes. The first is \textit{non-strategic agents}, i.e., agents always follow the recommendation. An underlying assumption of classical MAB algorithms, such behavior could be explicit in case the mechanism makes decisions for the agents; or implicit, e.g., agents are unaware of the entire set of alternatives or their desirability, or high cognitive overload is required to compute it. The second agent scheme is \textit{strategic agents}: the mechanism makes action recommendations, but cannot compel agents to follow these recommendations. In this scheme, we say that a mechanism is incentive compatible (IC) when following its recommendations is a dominant strategy: that is, when given a recommendation, an agent's best response is to follow her own recommendation. Formally,
\begin{definition}[Incentive Compatibility]\label{def:ic}
A mechanism $M$ is incentive compatible (IC) if $\forall l \in \{1,\ldots,n\}$, for every history
$h\in \left(A\times \R_+ \right)^{l-1}$ and for all actions $a_r,a_i \in A$,
\begin{equation}\label{eq: ic const}
\E(R^l(a_r)-R^l(a_i)\mid M(h)=a_r)\geq   0.
\end{equation}
\end{definition}
Unless stated otherwise, we address the non-strategic agents scheme. We handle the other agent scheme in Section \ref{sec:ic}.

When agents follow the mechanism, we can represent the mechanism's (expected) social welfare by
\begin{equation}
\label{eq: social welfare}
SW(M)=\E\left[ \frac{1}{n}\sum_{l=1}^n X_{M(h_l)} \right],
\end{equation}%
where $X_{M(h_l)}=\sum_{r=1}^K \Pr_{M(h_l)}\left(a_r\right) \E(X_{r} \mid h_l)$ is the reward agent $l$ receives. Notice that $X_{M(h_l)}$ depends on the randomness of the rewards \textit{and}, possibly, the randomness of $M(h_l)$.

Denote the highest possible social welfare under non-strategic agents by $\OPT$. A mechanism $M^*$ is said to be \textit{optimal} if $SW(M^*)= \OPT$. A mechanism $M^*$ is \textit{asymptotically optimal} if, for every ``large enough'' number of agents $n$ greater than some number $n'$, it holds that $SW(M^*) \geq \OPT- o(\frac{1}{n})$. 
This definition of approximation is equivalent to sub-linear regret in the MAB literature.

\subsection{Individual Guarantees}\label{subsec:individ guarantees}
An individual guarantee is a guarantee that a mechanism can provide to the agents it interacts with, independently of the other agents. In this subsection, we present our main conceptual contribution: a meaningful individual guarantee that allows exploration.  

To put our guarantee in the right context, we first present the strictest and the strongest guarantee that could be provided. A mechanism is a \textit{delegate} if it acts as the agent would have acted had it revealed the information it has with her. Formally, A mechanism $M$ is a delegate if for every agent $l\in\{1, \ldots, n\}$, every history $h \in \left(A\times \R_+ \right)^{l-1}$ and every distribution $\bl p$ over $A$, it holds that $
\E(X_{M(h)} \mid h)\geq \sum_{r=1}^K \bl p(r)\E(X_{r} \mid h)$. Indeed, this definition provides the strongest individual guarantee. It characterizes the greedy mechanism, \gre, which exploits in every round (according to the information it has). Noticeably, \gre performs little exploration, and probably leads to low social welfare. While sometimes relaxing this strong guarantee is impossible (e.g., banking or health-care), in many situations the planner is willing to relax individual guarantees to favor better social welfare.

The other extreme is to adopt a policy that we term \fulexp. \fulexp~ is the mechanism that first explores all arms sequentially, and then exploit the best arm. Clearly, at least for the non-strategic agent scheme, \fulexp~ is optimal when the number of agents is large enough. Nevertheless, with very high probability, it picks sub-optimal arms for the first $K$ agents, which can be a highly undesired property.

Our guarantee builds on the popular economic concept of individual rationality. To introduce it, we propose the following thought experiment. Assume that agents have to make decisions without the mechanism. The agents know that $\mu_1 =\max_i \mu_i$; hence, we shall assume that every agent's \textit{default action} is $a_1$.\footnote{As it will become apparent later, if agents have different default arms the social welfare can only increase since more arms could be explored.} The default action is the action each agent would have selected if she did not use the mechanism. We compare the two options: picking the default arm or following the mechanism's action. If a mechanism guarantees that the latter is higher in expectation according to its knowledge, agents are better off using the mechanism. As a result, an individually rational mechanism should guarantee each agent at least the reward obtained by her default action. The next definition relies on this reasoning. 
\begin{definition}[\textit{Ex-Ante} Individual Rationality]
A mechanism $M$ is \textit{ex-ante} individually rational (EAIR) if for every agent $l\in\{1, \ldots, n\}$, and every history $h \in \left(A\times \R_+ \right)^{l-1}$,
\begin{equation}
\label{eq: ir condition}
\sum_{r=1}^K \Pr\nolimits_{M(h)}(a_r) \E(X_{r} \mid h)  \geq \E(X_1 \mid h) .
\end{equation}
\end{definition}
The EAIR definition is conditioned on histories, i.e., the mechanism's knowledge. The right hand side is what an agent would get, given the knowledge of the mechanism, if she follows the default arm (which is optimal according to her knowledge). The left hand side is the expected value (over lotteries selected by the mechanism and reward distribution) guaranteed by the mechanism. Due to the mutual independence assumption, we must have $E(X_r\mid h)=R_r$ if arm $a_r$ was observed under the history $h$ and $E(X_r\mid h)=\mu_r$ otherwise. An EAIR mechanism must select a \textit{portfolio} of arms with expected reward never inferior to the reward of the default arm $a_1$.

\paragraph{Example.}
We now give an example to illustrate our setting and to familiarize the reader with our notation. Consider $K=3$ arms, $H=30$ and $X_1\sim Uni\{0,\dots 30\}, X_2\sim Uni\{0,\dots 20\}, X_3\sim Uni\{0,\dots 10\}$; 
thus $\mu_1=15$, $\mu_2=10$, and $ \mu_3=5$. As always, $a_1$ is the default arm. To satisfy EAIR, a mechanism should recommend $a_1$ to the first agent, since EAIR requires that the expected value of any recommendation should weakly exceed $R_1$. Let $h_1=(a_1,R_1)$ be the history after the first agent. Now, we have three different cases. First, if $R_1 > \mu_2=10$, we know that $\E(X_2 \mid h_1)<R_1$ and $\E(X_3 \mid h_1) <R_1$; therefore, an EAIR mechanism can never explore any other arm, since any distribution over $\{a_2,a_3\}$ would violate Inequality \eqref{eq: ir condition}. Second, if $R_1 \leq \mu_3 =5$, then $\E(X_2 \mid h_1) \geq R_1$ and $\E(X_3 \mid h_1) \geq R_1$, and hence an EAIR mechanism can explore both $a_2$ and $a_3$.

The third and most interesting case is where $\mu_3 < R_1 \leq \mu_2$, as when $R_1=8$. In this case, arm $a_3$ could only be recommended through a portfolio. An EAIR mechanism could select any distribution over $\{a_2,a_3\}$ that satisfies Inequality~\eqref{eq: ir condition}: any $p\in [0,1]$ such that $p\cdot \mu_2 +(1-p)\cdot \mu_3 \geq R_1$. This means that an EAIR mechanism can potentially explore arm $a_3$, yielding higher expected social welfare overall than simply recommending a non-inferior arm deterministically.

\section{Asymptotically Optimal EAIR Mechanism}\label{sec:aux mdp}
In this section, we consider the case of non-stratgic agents. We present the main technical contribution of this paper: a mechanism that asymptotically optimally balances the explore-exploit tradeoff while satisfying the EAIR property. The mechanism, which we term Fiduciary Explore \& Exploit (\mainalg), is described as Algorithm \ref{mainalg}. \mainalg is an event-based protocol that triggers every time an agent arrives. We now give an overview of \mainalg, focusing on the case where all agents adopt the recommendation of the mediator (we treat the other case in Section \ref{sec:ic}). 
We explain the algorithm's exploration phase in Subsection \ref{subsec:explore}, describe the overall algorithm in Subsection \ref{subsec: experience and exploit}, and prove the algorithm's formal guarantees in Subsection \ref{sec:algorithm}. We provide a comprehensive example of the way \mainalg operates in~{\ifnum\Includeappendix=0{the appendix.
}\else{Section \ref{sec:elaborated example.}.}\fi}

\mainalg is composed of three phases: primary exploration (Lines \ref{alg:initialize}--\ref{alg:update s}), secondary exploration (Lines \ref{alg:else towards experience}--\ref{alg:reuse good arm}), and exploitation (Lines \ref{alg: best arm}).
During the primary exploration phase, the mechanism compares the default arm $a_1$ to whichever other arms are permitted by the individual rationality constraint. This turns out to be challenging for two reasons. First, the order in which arms are explored matters; tackling them in the wrong order can reduce the set of arms that can be considered overall. Second, it is nontrivial to search in the continuous space of probability distributions over arms. To address this latter issue, we present a key lemma that allows us to use dynamic programming and find the optimal exploration policy in time $O(2^K K^2 H^2 )$. Because we expect $K$ either to be fixed or to be significantly smaller than $n,H$, this policy is computationally efficient. Moreover, we note that the optimal exploration policy can be computed offline prior to the agents' arrival.

The primary exploration phase terminates in one of two scenarios: either the reward $R_1$ of arm $a_1$ is the best that was observed and thus no other arm could be explored (as in our example when $R_1 > 10$, or when $R_1=8$ and exploring $a_2$ yielded $R_2 \leq R_1$ and thus $a_3$ could not be explored), or another arm $a_i$ was found to be superior to $a_1$: i.e., an arm $a_i$ was observed for which $R_i>R_1$. In the latter case, the mechanism gains the option of conducting a secondary exploration, using arm $a_i$ to investigate all the arms that were not explored in the primary exploration phase. The third and final phase---to which we must proceed directly after the primary exploration phase if that phase does not identify an arm superior to the default arm---is to exploit the most rewarding arm observed. 

\textbf{Remark.} In this section we assume that agents are non-strategic and follow the mechanism's recommendation.

\subsection{Primary Exploration Phase}
\label{subsec:explore}
Performing primary exploration optimally requires solving a planning problem; it is a challenging one, because it involves a continuous action space and a number of states exponential in $K$ and $H$. We approach this task as a Goal Markov Decision Process (GMDP) (see, e.g., \cite{barto1995learning}) that abstracts everything but pure exploration. In our GMDP encoding, all terminal states fall into one of two categories. The first category is histories that lead to pure exploitation of $a_1$, which can arise either because EAIR permits no arm to be explored or because all explored arms yield rewards inferior to the observed $R_1$; the second is histories in which an arm superior to $a_1$ was found. Non-terminal states thus represent histories in which it is still permissible for some arms to be explored. The set of actions in each non-terminal state is the set of distributions over the non-observed arms (i.e., portfolios) corresponding to the history represented in that state, which satisfy the EAIR condition. The transition probabilities encode the probability of choosing each candidate arm from a portfolio; observe that the rewards of each arm are fixed, so this is not a source of additional randomness in our model. GMDP rewards are given in terminal nodes only: either the observed $R_1$ if no superior arm was found or the expected value of the maximum between the superior reward discovered and the maximal reward of all unobserved arms (since in this case, as we show later on, the mechanism is able to explore all arms w.h.p. during the secondary exploration phase).

Formally, the GMDP is a tuple $\langle \mS, \mA, \mP, \mR \rangle$, where
\begin{itemize}[leftmargin=*,wide=\parindent]
\item $\mS$ is a finite set of states. Each state $s$ is a pair $(O,U)$, where $O \subseteq \{(a,c)\mid a\in A, c\in H \}$ is the set of arm-reward pairs that have been observed so far, with each $a$ appearing at most once in $O$ (since rewards from the arms are deterministic): for every $(O,U)$ and every $a\in A$, $\abs{\{c\mid (a,c)\in O  \} }\leq 1$. $U\subseteq A$ is the set of arms not yet explored. The initial state is thus $s_0=(\emptyset, A)$. For every non-empty\footnote{Due to the construction, every non-empty $O$ must contain $(a_1,c)$ for some $c\in [H]^+$.} set of pairs $O$ we define $\alpha(O)$ to be the reward observed for arm $a_1$, and $\beta(O)= \max_{c:\exists a, (a,c)\in O}c$ to be the maximal reward observed.

\item $\mA=\bigcup_{s \in \mS} \mA_s$ is an infinite set of actions. For each $s=(O,U)\in \mS$, $\mA_s$ is defined as follows:
	\begin{enumerate}
	\item If $s=s_0$, then $\mA_{s_0}=\Delta(\{a_1\})$: i.e., a deterministic selection of $a_1$.
	\item Else, if $\alpha(O) < \beta(O) $, then $\mA_s=\emptyset$. This condition implies that we can move to secondary exploration.
	\item Otherwise, $\mA_s$ is a subset of $\Delta(U)$, such that $\bl p \in \mA_s$ if and only if {\ifnum\Includeappendix=0{$
\sum_{a_i\in U} \bl p(a_i) \mu_{a_i} \geq \alpha(O).$
    }\fi}
	{\ifnum\Includeappendix=1{\begin{equation}\label{eq:gmdp EAIR}
\sum_{a_i\in U} \bl p(a_i) \mu_{a_i} \geq \alpha(O).
\end{equation}}\fi}Notice that this resembles the EAIR condition given in Inequality~\eqref{eq: ir condition}. Moreover, the case where none of the remaining arms have strong enough priors to allow exploration falls here as a vacuous case of the above inequality.
\end{enumerate}
We denote by $\mS_T$ the set of \textit{terminal} states, namely $\mS_T =\{s\in\mS \mid \mA_s =\emptyset \}$.

\item $\mP$ is the transition probability function. Let $s=(O,U) \in \mS$, and let $s'=(O',U')$ such that $O' =O\cup \{(a_i,c)\}$ and $U'=U\setminus\{a_i\}$ for some $a_i\in U, c\in [H]^+$. Then, the transition probability from $s$ to $s'$ given an action $\bl p$ is defined by
$
\mP(s'|s,\bl p)= \bl p (a_i) \Pr(X_i=c) .
$
If $s'$ is some other state that does not meet the conditions above, then let $\mP(s'|s,\bl p)=0$ for every $\bl p \in \mA_s$.
\item $\mR : \mS_T \rightarrow \mathbb{R}$ is the reward function, defined on terminal states only. For each terminal state $s=(O,U) \in \mS_T$,
{\thinmuskip=0mu
\medmuskip=0mu plus 0mu minus 0mu
\thickmuskip=0mu plus 0mu
\[
\mR(s)=
\begin{cases}
\alpha(O) &  \alpha(O)=\beta(O)\\
\E\left[\max\left\{\beta(O), \max_{a_{i'}\in U}X_{i'}) \right\} \right]  & \alpha(O)<\beta(O)
\end{cases}.
\]}%
That is, when $a_1$ was the highest-reward arm observed, the reward of a terminal state is $\alpha(O)$; otherwise, it is the expectation of the maximum between $\beta(O)$ and the highest reward of all unobserved arms. The reward depends on unobserved arms since the secondary exploration phase allows us to explore all these arms; hence, their values are also taken into account. 
\end{itemize}
A policy $\pi: (\mS \times \mA)^* \times \mS \rightarrow \mA$ is a function from all GMDP histories (sequences of states and actions) and a current state to an action. A policy $\pi$ is \textit{valid} if for every history $h$ and every non-terminal state $s$, $\pi(h,s)\in \mA_s$. A policy $\pi$ is \textit{stationary} if for every two histories $h,h'$ and a state $s$, $\pi(h,s)=\pi(h',s)$. When discussing a stationary policy, we thus neglect its dependency on $h$, writing $\pi(s)$.

Given a policy  $\pi$ and a state $s$, we denote by $W(\pi,s)$ the expected reward of $\pi$ when initialized from $s$, which is defined recursively from the terminal states:
{\thinmuskip=0mu
\medmuskip=0mu plus 0mu minus 0mu
\thickmuskip=0mu plus 0mu
\begin{equation*}
W(\pi,s) =
\begin{cases}
\mR(s) & \text{if } s\in S_T \\
\sum_{s'\in \mS} \mP(s'|s,\pi(s))W(\pi,s') & \text{otherwise.}
\end{cases}
\end{equation*}}%
We now turn to our technical results. The following lemma shows that we can safely focus on stationary policies that effectively operate on a significantly reduced state space.  
\begin{lemma}\label{lemma:gmdp reduction}
For every policy $\pi$ there exists a stationary policy $\pi'$ such that (1) $\pi'(s)=\pi'(s')$ for every pair of states $s=(O,U)$ and $s'=(O',U)$ with $\alpha(O)=\alpha(O')$ and  $\beta(O)=\beta(O')$; and (2)
 for every state s,  $W(\pi',s)\geq W(\pi,s)$.
\end{lemma}
Lemma \ref{lemma:gmdp reduction} tells us that there exists an optimal, stationary policy that selects the same action in every pair of states that share the same unobserved set $U$ and values $\alpha(O)$ and $\beta(O)$, but are distinguished in the $O$ component. Thus, we do not need a set of states whose size depends on the number of possible arm-reward observation histories: all we need to record is $U$ and a real value for either $\alpha(O)$ and $\beta(O)$, reducing the number of states to $O(2^K H)$.

We still have one more challenge to overcome: the set of actions $\mA_s$ available in each state $s$ is infinite. Despite that $\mA_s$ is a convex polytope and thus we can apply Linear Programming, our approach is much more computationally efficient and interpretable. We prove that there exists an optimal ``simple'' policy, which we denote $\pi^*$. Given two indices $i,r \in \{2,\dots,K\}$, we denote by $\bl p^\alpha_{ir}$ (for $i\neq r$) and by $\bl p^\alpha_{ii}$ (for $i=r$) the distributions over $\{a_1,\dots,a_K\}$ such that  
{\thinmuskip=0mu
\medmuskip=0mu plus 0mu minus 0mu
\thickmuskip=0mu plus 0mu
\[
\bl p^\alpha_{ir}(a) =
\begin{cases}
\frac{\abs{\alpha-\mu_r}}{\abs{\alpha-\mu_i}+\abs{\alpha-\mu_r}} &\text{if } a=a_i \\
\frac{\abs{\alpha-\mu_i}}{\abs{\alpha-\mu_i}+\abs{\alpha-\mu_r}} &\text{if } a=a_r \\
0 &\text{otherwise}
\end{cases},
\]}%
and $\bl p^{\alpha}_{ii}(a)=1$ if and only if $a=a_i$. When $\alpha=\alpha(O)$ is clear from context, we omit it from the superscript. 

We are now ready to describe the policy $\pi^*$, which we later prove to be optimal. For the initial state $s_0$, $\pi^*(s_0)=\bl p_{11}$. For every non-terminal state $s=(O,U)\in\mS$ with $s\neq s_0$, $\pi^*(s)= \bl p_{i^*r^*}$ such that $(i^*,r^*)\in \mA_s$ maximize
{\thinmuskip=0mu
\medmuskip=0mu plus 0mu minus 0mu
\thickmuskip=0mu plus 0mu
\begin{align*}
&\left(
1-\frac{\ind_{i=r}}{2}
\right)
\bigg[
\bl p_{ir}(i)\sum_{s'\in \mS} \mP(s'|s,\bl p_{ii})W(\pi^*,s')\\
&\qquad \qquad\qquad +\bl p_{ir}(r)\sum_{s'\in \mS} \mP(s'|s,\bl p_{rr})W(\pi^*,s')
\bigg].
\end{align*}
}%
The optimality of $\pi^*$ follows from a property that is formally proven in Theorem \ref{thm:optimal policy}: any policy $\pi$ that satisfies the conditions of Lemma \ref{lemma:gmdp reduction} can be presented as a mixture of policies that solely take actions of the form $(\bl p_{ir})_{i,r}$. As a result, we can improve $\pi$ by taking the best such policy from that mixture.  We derive $\pi^*$ via  dynamic programming, where the base cases are the set of terminal states. For any other state, $\pi^*(s)$ is the best action of the form $\bl p_{ir}$ as defined above, considering all states that are reachable from $s$. While any policy $\pi'$ can be encoded as a weighted sum over such ``simple'' policies, $\pi^*$ is the best one, and hence is optimal.
\begin{theorem}\label{thm:optimal policy}
For every valid policy $\pi$ and every state $s$, it holds that $W(\pi^*,s) \geq W(\pi,s)$.
\end{theorem}
Since our compressed state representation consists of $O(2^K H)$ states, the computation of $\pi^*$ in each stage requires us to consider $O(K^2)$ candidate actions, each of which involves summation of at most $H+1$ summands; thus, $\pi^*$ can be computed in $O(2^K K^2 H^2)$ time.

\subsection{Intuitive Description of \mainalg}\label{subsec: experience and exploit}

We now present the \mainalg algorithm, stated formally as Algorithm \ref{mainalg}. The primary exploration phase (Lines \ref{alg:initialize}--\ref{alg:update s}) is based on the GMDP from the previous subsection. It is composed of computing $\pi^*$ and then producing recommendations according to its actions, each of which defines a distribution over (at most) two actions. Let $(U,O)$ denote the terminal state reached by $\pi^*$ (the primary exploration selects a fresh arm in each stage; hence such a state is reached after at most $K$ agents).

We then enter the secondary exploration phase. If $\beta(O)= R_1$ then this phase is vacuous: no distribution over the unobserved arms can satisfy the EAIR condition and/or all the observed arms are inferior to arm $a_1$. On the other hand, if $\beta(O) > R_1$ (Line \ref{alg:else towards experience}), we found an arm $a_{\tilde{r}}$ with a reward superior to $R_1$, and can use it to explore all the remaining arms. For every $a_i \in U$, the mechanism operates as follows. If the probability of $a_i$ yielding a reward greater than $a_{\tilde{r}}$ is zero, we neglect it (Lines \ref{alg: dominated}--\ref{alg: while explore dominated continue}). Else, if $\mu_i \geq R_1$, we recommend $a_i$. This is manifested in the second condition in Line \ref{alg:while if conditions}. Otherwise, $\mu_i<R_1$. In this case, we select a distribution over $\{a_{\tilde{r}},a_i\}$ that satisfies the EAIR condition and explore $a_i$ with the maximal possible probability, which is $\bl p_{\tilde r i}(i)$. As we show formally in the proof of Lemma~\ref{lemma:our alg sw is best}, the probability of exploring $a_i$ in this case is at least $\frac{1}{H}$, implying that after $H$ tries in expectation the algorithm would succeed to explore $a_i$.

Ultimately (Line \ref{alg: best arm}), \mainalg recommends the best observed arm to all the remaining agents.

\subsection{Algorithmic Guarantees}\label{sec:algorithm}
\begin{algorithm}[t]
\caption{Fiduciary Explore \& Exploit \label{mainalg}(FEE)}
\begin{algorithmic}[1]
\STATE Initialize a GMDP instance $\langle \mS, \mA, \mP, \mR \rangle$, and compute $\pi^*$.\label{alg:initialize}
\STATE Set $s=(O,U)=(\emptyset,A)$. 
\WHILE {$s$ is not terminal\label{alg:while not terminal}}{
\STATE Draw arm $a_i\sim \pi^*(s)$, recommend $a_i$ and observe $R_i$.\label{alg:using policy}
\STATE $O\gets O \cup \{(a_i,R_i)\}, U \gets U\setminus \{a_i\}$.
\STATE $s\gets (O,U)$.\label{alg:update s}
}
\ENDWHILE
\IF{$\beta(O)>R_1$  \label{alg:else towards experience}}
{
\WHILE%
{$U$ is not empty \label{alg:while loop}}{
\STATE Let $a_{\tilde r}$ s.t. $a_{ \tilde r} \in \argmax_{a_r \in A\setminus U} R_{r}$. \label{alg: pick a r tilde}
\STATE Select an arbitrary arm $a_i\in U$. \label{alg:select a_i}
\IF{$\Pr(X_i > R_{\tilde r})=0 \label{alg: dominated} $}
{
\STATE $U \gets U\setminus \{a_i\}$.  \label{alg: while explore dominated}
\STATE\textbf{continue}. \label{alg: while explore dominated continue}
}
\ENDIF
\STATE Draw $Y\sim Uni[0,1]$. \label{alg:draw} 
\IF{$Y \leq \frac{R_{\tilde r}-R_1}{R_{\tilde r}-\mu_i}$ or $\mu_i \geq R_1$ \label{alg:while if conditions}}
 {
\STATE Recommend $a_i$ and observe $R_i$. \label{alg:discover new arm}
\STATE $U \gets U\setminus \{a_i\}$. \label{alg: while explore open}
}
\ENDIF
\STATE \lELSE {recommend $a_{\tilde r}$. \label{alg:reuse good arm} }
}
\ENDWHILE
}
\ENDIF
\STATE Recommend $a_{i^*}\in \argmax_{a_i \in A \setminus U} R_i$ to all agents. \label{alg: best arm}
\end{algorithmic}
\end{algorithm}
We begin by arguing that \mainalg is indeed EAIR.
\begin{proposition}\label{prop:is EAIR}
\textnormal{\mainalg} satisfies the EAIR condition.
\end{proposition}
The proof of Proposition~\ref{prop:is EAIR} is highly intuitive: the reward of every recommendation {\mainalg} makes always exceed $R_1$ in expectation. We now move on to consider the social welfare of \mainalg. Let $\OPTEA$ denote the highest welfare attained by any EAIR mechanism. First, we show that the expected value of $\pi^*$ at $s_0$, denoted by $W(\pi^*,s_0)$, upper bounds the social welfare of any EAIR mechanism.
\begin{theorem} \label{theorem:ic-ir upper}
It holds that $\OPTEA\leq W(\pi^*,s_0)$.
\end{theorem}
The proof proceeds by contradiction: given an EAIR mechanism $M$, we construct a series of progressively-easier-to-analyze EAIR mechanisms with non-decreasing social welfare; we modify the final mechanism by granting it oracular capabilities, making it violate the EAIR property and yet preserving reducibility to a policy for the GMDP of Subsection \ref{subsec:explore}. We then argue via the optimality of $\pi^*$ that the oracle mechanism cannot obtain a social welfare greater than $W(\pi^*,s_0)$. Next, we lower bound the social welfare of \mainalg.
\begin{lemma} \label{lemma:our alg sw is best} 
$
SW_{n}(\textnormal{\mainalg}) \geq \OPTEA-O\left(\frac{KH^2}{n}  \right).
$
\end{lemma}
The proof relies mainly on an argument that the primary and secondary explorations will not be too long on average: after $(K+1)H$ agents the mechanism is likely to begin exploiting. Noting that the lower bound of Lemma \ref{lemma:our alg sw is best} asymptotically approaches the upper bound of Theorem \ref{theorem:ic-ir upper}, we conclude that \mainalg is  asymptotically optimal.

\section{Incentive Compatibility}\label{sec:ic}
In this section, we consider the second and more challenging agent scheme: strategic agents. Our main goal is to show that \mainalg, which we developed in Section \ref{sec:algorithm}, can be modified to satisfy IC as well.\footnote{
For simplicity, we formulated \mainicalg to satisfy IC in the best response sense: given that all other agents follow their recommendations, it is an agent's best response to adopt the recommendation as well. However, \mainicalg can be easily modified to offer dominant strategies to agents.}
We remark that there are cases that an IC mechanism cannot explore all arms, regardless of individual rationality constraints. To illustrate, assume that $\Pr(X_1\geq \mu_2)=1$, i.e., the reward of arm $a_1$ is always greater or equal to the expected reward of arm $a_2$. In this case, no agent will ever follow a recommendation for arm $a_2$. Consequently, we shall make the following standard assumption (see, e.g., \cite{Mansour2015})
\begin{assumption}\label{assumption ic}
For every $i,j$ such that $1\leq i<j \leq K$, it holds that $\Pr(X_i < \mu_j)>0$.
\end{assumption}
If Assumption \ref{assumption ic} does not hold for some pair $(i,j)$, arm $a_j$ would never be explored; hence, we can remove such arms from $A$. We shall also make the simplifying assumption that $\mu_1>\mu_2 \geq \dots \geq \mu_K$, as otherwise the problem becomes easier to solve.

Among other factors, the expectation in Inequality (\ref{eq: ic const}) is taken over agents' information on the arrival order. On the one extreme, the arrival order could be uniform, i.e., each agent $l$ is entirely oblivious about her "place in line." 
In this case, as we show in~{\ifnum\Includeappendix=0{the appendix,}\else{Section \ref{sec:ic and uniform},}\fi} \mainalg satisfies IC as is assuming that there are sufficiently many agents. On the other extreme, which is the more popular in prior work \cite{Kremer2014,Mansour2015}, agents have complete information about their rounds. Namely, the agent arriving at time $l$ knows that she is the $l$'th agent. The complete information case is the more demanding one, and an IC mechanism for this case will also be IC under any distributional assumption on the arrival order. Nevertheless, as we demonstrate shortly, it requires more technical work.

We build on the techniques of \citet{Mansour2015} and use \textit{phases}: each phase contains one round of exploration (that is, following \mainalg) and the other rounds are either exploitation via \gre (defined in Subsection \ref{subsec:individ guarantees}) or recommendation of arm $a_1$. An IC version of \mainalg, which we term \mainicalg, is outlined as Algorithm \ref{mainicalg}. 

\mainicalg works as follows. It initializes an instance of \mainalg, and uses it seldom in the earlier rounds, and regularly afterward (every time \mainicalg makes a recommendation, it updates \mainalg). In Line \ref{alg2:first agent}, it recommends $a_1$ to the first agent. Recall that $\pi^*$ employed by \mainalg is only allowed to pick $a_1$ w.p. 1 in the first round; hence,  \mainalg and \mainicalg coincide with the first recommendation. Then, depending on the value of $R_1$, it recommends agents $2,\dots,K$ either greedily (maximizing the reward in each round, Line \ref{alg2: r1 is bad}) or arm $a_1$ (Line \ref{alg2: r1 after one}). Later, in Line \ref{alg2:split}, it splits the remaining rounds to \textit{phases} of size $B$ ($B$ will be determined later on). In each such phase $k$, we first ask whether \mainalg is exploring or exploiting (Line \ref{alg2:if exploits}). If \mainalg exploits (Line \ref{alg: best arm} in Algorithm \ref{mainalg}), every agent of every phase from here on will be recommended by \mainalg. If that is not the case (see the else block starting at Line \ref{alg2:else block explore}), \mainicalg picks one agent from the $B$ agents of this phase uniformly at random, denoted $l(k)$. Then, agent $l(k)$ gets the recommendation from \mainalg. The recommendation policy for the rest of the agents in this phase depends on the observed arms. If \mainicalg already discovered an arm $a_i$ with $R_i>R_1$ (Line \ref{alg2: recommend as greedy if observed}), we let agent $l$ exploit using \gre. Otherwise (Line \ref{alg2: recommend a on in phase}), \mainicalg recommends $a_1$. 

Lines \ref{alg2: recommend as greedy if observed} and \ref{alg2: recommend a on in phase} are also where our mechanism departs from the principles of prior work. For example, in the work of \citet{Mansour2015}, each phase contains one round of exploration and the rest are exploitation rounds. In our setting, agents that are not exploring might still not exploit. 
The distinction between Lines \ref{alg2: recommend as greedy if observed} and \ref{alg2: recommend a on in phase} is crucial: exploiting unobserved arms might lead to sub-optimal welfare, since they are the chance to explore arms with expected reward below $R_1$. We elaborate more in the proof of Theorem \ref{theorem: ic fee}.

To determine the phase length $B$, we introduce the following quantities $\xi$ and $\gamma$. Due to Assumption \ref{assumption ic}, there exist $\xi>0$ and $\gamma>0$ such that for all $i\in [K]$, it holds that $\Pr(\forall i'\in [K]\setminus \{i\}:\mu_i -X_{i'} > \xi  )>\gamma$. In words, it says that the reward of every arm $i$ is greater than all other arms by at least $\xi$, w.p. of at least $\gamma$. The following Theorem \ref{theorem: ic fee} summarizes the properties of \mainicalg.
\begin{theorem}\label{theorem: ic fee}
Let the phase length be $B= \ceil*{\frac{H}{\xi \gamma}}+1$. Under Assumption \ref{assumption ic}, $\textnormal{\mainicalg}$ satisfies EAIR and IC. In addition,
$
SW_{n}(\textnormal{\mainicalg}) \geq \OPTEA-O\left(\frac{KH^3}{n\xi \gamma}  \right).
$
\end{theorem}

\begin{algorithm}[t]
\caption{IC Fiduciary Explore \& Exploit \label{mainicalg}(IC-FEE)}
\begin{algorithmic}[1]
\STATE Initialize an instance of \mainalg and update it after every recommendation.\label{alg2:initialize} 
\STATE Recommend $a_1$ to the first agent, observe $R_1$. \label{alg2:first agent}
\STATE\lIF{$R_1 < \mu_K$}{recommend as \gre to agents $2,\dots, K$.\label{alg2: r1 is bad}}%
\STATE\lELSE{recommend $a_1$ to agents $2,\dots, K$. \label{alg2: r1 after one}
}
\STATE Split the remaining rounds into consecutive phases of $B$ rounds each. \label{alg2:split}
\FOR {phase $k=1,\dots$ }{
\IF{\mainalg exploits (Line \ref{alg: best arm} in \mainalg) \label{alg2:if exploits}}
 {
\STATE follow \mainalg \label{alg2:exploits}. 
}
\ELSE {\label{alg2:else block explore}
\STATE Pick one agent $l(k)$ from the $B$ agents in this phase uniformly at random, and recommend her according to \mainalg. \label{alg2: pick to explore}

As for the rest of the agents,
\STATE \lIF{an arm $a_i$ with $R_i>R_1$ was revealed}{recommend as \gre.} \label{alg2: recommend as greedy if observed}
\STATE \lELSE{recommend $a_1$.} \label{alg2: recommend a on in phase}
}
\ENDIF
}
\ENDFOR
\end{algorithmic}
\end{algorithm}

\section{Further Analysis}\label{sec:analysis}
Notice that EAIR mechanisms guarantee each agent the value of the default arm, but only in expectation. We now propose a more strict form of individual rationality, \textit{ex-post} individual rationality (EPIR).
\begin{definition}[\textit{Ex-Post} Individual Rationality]
A mechanism $M$ is \textit{ex-post}  individually rational (EPIR) if for every agent $l\in\{1, \ldots, n\}$, every history $h\in \left(A\times \R_+ \right)^{l-1}$, and every arm $a_r$ such that $\Pr_{M(h)}(a_r)>0$, it holds that $
\E(X_{r}-X_1 \mid h)  \geq 0.$
\end{definition}
Satisfying EPIR means that the mechanism never recommends an arm that is \textit{a priori} inferior to arm $a_1$ given the mechanism's knowledge. It is immediate to see that every EPIR mechanism is also EAIR. EPIR mechanisms are quite conservative, since they can only explore arms that yield expected rewards of at least the value $R_1$ obtained for $a_1$. We develop an optimal IC/EPIR mechanism in~{\ifnum\Includeappendix=0{the appendix}\else{Section \ref{sec:EPIR}}\fi}.

\subsection{Social Welfare Analysis}\label{subsec:welfare analysis}
We now analyze the loss in social welfare due to  individual rationality constraints. For simplicity, we consider the case of non-strategic agents. Recall that $\OPT$ is the highest possible social welfare, and $\OPTEA$ is its counterpart after imposing EAIR. In addition, let $\OPTEP$ and $\OPTDEL$ denote the best asymptotic social welfare (w.r.t.\ some instance $\langle K,A,(X_i) \rangle$ and infinitely many agents) achievable by an EPIR and a delegate mechanisms, respectively. Noticeably, for every instance $\langle K,A,(X_i) \rangle$, it holds that $\OPT\geq\OPTEA\geq\OPTEP\geq \OPTDEL$.
In the rest of this subsection, we analyze the ratio of two subsequent optimal welfares. We begin by showing that individual guarantees can deteriorate welfare even for the most flexible notion, EAIR. 
\begin{proposition} \label{prop:bound on EAIR}
For every $K,H\in \mathbb N$, there exists an instance $\langle K,A,(X_i) \rangle$ with $\frac{\OPT}{\OPTEA} \geq H\left(1-e^{-\frac{K}{H}}\right)$. 
\end{proposition}
Proposition \ref{prop:bound on EAIR} shows that when $K$ and $H$ have the same magnitude, the ratio is on the order of $H$, meaning that EAIR mechanisms perform poorly when a large number of different reward values are possible. However, this result describes the worst case; it turns out that optimal EAIR mechanisms have constant ratio under some reward distributions. For example, as we show in~{\ifnum\Includeappendix=0{the appendix}\else{Proposition \ref{prop:bound on EAIR uniform}}\fi} this ratio is at most $\frac{8}{7}$ if $X_i \sim Uni\{0,1,\dots H\}$ for every $i\in\{2,\dots,K\}$ and $X_1$ is only slightly better a-priori. 

Next, we consider the cost of adopting the stricter EPIR condition rather than EAIR. As Proposition~\ref{prop:bound on EAIR-EPIR} shows, by providing a more strict fiduciary guarantee the social welfare may be harmed by a factor of $H$. 
\begin{proposition} \label{prop:bound on EAIR-EPIR}
For every $K,H\in \mathbb N$, there exists an instance $\langle K,A,(X_i) \rangle$ with $\frac{\OPTEA}{\OPTEP} \geq \frac{H+2}{3}\left(1-e^{-\frac{K-2}{H}} \right)$.
\end{proposition}
Finally, we show that the EPIR guarantee still allows us to significantly improve upon $\OPTDEL$.
\begin{proposition}
\label{prop:bound on EPIR-del}
For every $K,H\in \mathbb N$, there exists an instance $\langle K,A,(X_i) \rangle$ with $\frac{\OPTEP}{\OPTDEL} \geq \frac{H}{3}\left( 1-e^{-\frac{K-2}{H}}\right)$.
\end{proposition}

\section{Conclusions and Discussion}\label{discussions}

This paper introduces a model in which a recommender system must manage an exploration-exploitation tradeoff under the constraint that it may never knowingly make a recommendation that will yield lower reward than any individual agent would achieve if he/she acted without relying on the system. 

We see considerable scope for follow-up work. First, from a technical point of view, our algorithmic results are limited to discrete reward distributions. One possible future direction would be to present an algorithm for the continuous case. More conceptually, we see natural extensions of EPIR and EAIR to stochastic settings, either by assuming a prior and requiring the conditions w.r.t.\ the posterior distribution or by requiring the conditions to hold with high probability. Moreover, we are intrigued by non-stationary settings---where e.g., rewards follow a Markov process---since the planner would be able to sample \emph{a priori} inferior arms with high probability assuming the rewards change fast enough, thereby reducing regret.

\section*{Acknowledgements}
We thank the participants of the Computational Data Science seminar at Technion -- Israel Institute of Technology and the participants of Young Researcher Workshop on Economics and Computation for their comments and suggestions. Additionally, we thank ICML 2020 anonymous reviewers who provided comments that improved the manuscript. The work of G. Bahar, O. Ben-Porat and M. Tennenholtz is funded by the European Research Council (ERC) under the European Union's Horizon 2020 research and innovation programme (grant agreement n$\degree$  740435). The work of K. Leyton-Brown is funded by the NSERC Discovery Grants program, DND/NSERC Discovery Grant Supplement, Facebook Research and Canada CIFAR AI Chair Amii. Part of this work was done while K. Leyton-Brown was a visiting researcher at Technion -- Israel Institute of Science and was partially funded by the European Union's Horizon 2020 research and innovation programme (grant agreement~n$\degree$~740435).

\bibliographystyle{abbrvnat} 

{\ifnum\Includeappendix=1{ 
\newpage
\onecolumn
\appendix
\section{Omitted Proofs from Subsection \ref{subsec:explore}}

\begin{proofof}{Lemma \ref{lemma:gmdp reduction}}
The proof follows from Propositions \ref{prop:stationary is enough} and \ref{prop o oblivious} below.
\end{proofof}

\begin{proposition}\label{prop:stationary is enough}
For every non-stationary policy $\pi$, there exists a stationary policy $\pi'$ such that for every state $s\in \mS$, $W(\pi,s) \leq W(\pi',s)$.
\end{proposition}
Moreover, the following Proposition \ref{prop o oblivious} implies that we can substantially reduce the state space by disregarding the observed part $O$ and 
\begin{proposition}\label{prop o oblivious}
For every stationary policy $\pi$ there exists a stationary policy $\pi'$ such that:
\begin{enumerate}
\item  $\pi'(s)=\pi'(s')$ for every pair of states $s=(O,U),s'=(O',U)$ with $\alpha(O)=\alpha(O')$ and $\beta(O)=\beta(O')$.
\item for every state s,  $W(\pi',s)\geq W(\pi,s)$.
\end{enumerate}
\end{proposition}

\begin{proofof}{Proposition \ref{prop:stationary is enough}}
Fix an arbitrary non-stationary policy $\pi$. We prove the claim by iterating over all states in an increasing order of the number of elements in $U$. We use induction to show that the constructed $\pi'$ indeed satisfies the assertion. 

For every $s=(O,U)\in S$ such that $\abs{U}=1$, i.e., $U=\{a\}$. If $s$ is terminal, then $\mA_s=\emptyset$ and $W(\pi,s) = W(\pi',s) = \alpha(O)$. Otherwise, the unique element in $\mA_s$ is the action that assigns probability 1 to $a$, and by setting $\pi'(s)=\pi(s)$ we get $W(\pi',s)= W(\pi,s)$.

Assume that the assertion holds for every $\abs{U}\leq j$; namely, that $W(\pi',s)\geq W(\pi,s)$ for all $s= (O,U)\in \mS$ with $\abs{U}\leq j$. We now prove the assertion for $s= (O,U)\in \mS$ with $\abs{U}= j+1$. If $s$ is a terminal state, then we are done. Else, since $U$ and the support of each arm are finite, there exists a finite number of possible histories that lead from $s_0$ to $s$ that we will mark as $h_1, \ldots h_w$. For every possible history $h\in \{h_1, \ldots h_w \} $, $\pi$ assigns an action $\bl p_{h} \in \mA_s$ that (can) depend on the history $h$. Let
\begin{align}
\small
\bl p^* \in \argmax_{\bl p_h, h\in \{h_1,\ldots ,h_w\}} 
\Bigg\{
&\sum_{a_i\in U}\bl p_{h}(a_i)\sum_{s'\in \mS}\mP(s'\mid s,\bl p_{ii})W(\pi,s') 
\Bigg\},
\end{align} 
breaking ties arbitrarily. We set $\pi'(s)= \bl p^*$ . Hence we get:
\begin{align*}
W(\pi,s)&=\sum_{s'\in \mS,h\in \{h_1, \ldots h_w\} } \Pr(h)\mP(s'\mid s,\pi, h)W(\pi,s') \\
&=\sum_{a_i\in U, h\in \{h_1, \ldots h_w\}}\Pr(h)\bl p_{h}(a_i)\sum_{s'\in \mS}\mP(s'\mid s,\bl p_{ii})W(\pi,s') \\
&\leq \sum_{h\in \{h_1, \ldots h_w\}}\Pr(h)\argmax_{h\in \{h_1,\ldots ,h_w\}} 
\Bigg\{
\bigg(
\sum_{a_i\in U}\bl p_{h}(a_i)\sum_{s'\in \mS}\mP(s'\mid s,\bl p_{ii})W(\pi,s') 
\bigg)
\Bigg\} \\
&=1\argmax_{h\in \{h_1,\ldots ,h_w\}} 
\Bigg\{
\sum_{a_i\in U}\bl p_{h}(a_i)\sum_{s'\in \mS}\mP(s'\mid s,\bl p_{ii})W(\pi,s') 
\Bigg\} \\
&=\argmax_{h\in \{h_1,\ldots ,h_w\}} 
\Bigg\{
\sum_{a_i\in U}\bl p_{h}(a_i)\sum_{s'\in \mS}\mP(s'\mid s,\bl p_{ii})W(\pi',s') 
\Bigg\} \\
&=\sum_{a_i\in U}\bl p^*(a_i)\sum_{s'\in \mS}\mP(s'\mid s,\bl p_{ii})W(\pi',s') \\
&= W(\pi',s);
\end{align*}
hence, $W(\pi,s)\leq W(\pi',s)$. This concludes the proof.
\end{proofof}

\begin{proofof}{Proposition \ref{prop o oblivious}}
The proof is similar to the proof of Proposition \ref{prop:stationary is enough}, and is given for completeness. Fix an arbitrary stationary policy $\pi$. We prove the claim by iterating over all states in an increasing order of the number of elements in $U$. We use induction to show that the constructed $\pi'$ indeed satisfies the assertion. 

For every $s=(O,U)\in S$ such that $\abs{U}=1$, i.e., $U=\{a\}$, if $s$ is terminal then $W(\pi,s) = W(\pi',s) = \alpha(O)$. Otherwise, the unique element in $\mA_s$ is the action that assigns probability 1 to $a$; hence, by setting $\pi'(s)=\pi(s)$ we get $W(\pi',s)= W(\pi,s)$.

Assume the assertion holds for every $\abs{U}\leq j$; namely, that $W(\pi',s)\geq W(\pi,s)$ for all $s= (O,U)\in \mS$ with $\abs{U}\leq j$. Next, we prove the assertion for $s= (O,U)\in \mS$ with $\abs{U}= j+1$. If $s$ is a terminal state, then we are done. Else, since the size of $O$ and the support of each arm are finite, there exists only a finite number of states with the same $U$ and $\alpha$, which we mark as $s = s_0=(O,U) , s_1 = (O^1,U), \ldots s_w = (O^w,U)$. For every state $s_j = (O^j, U)$, $\pi$ assigns an action $\bl p_{s_j} \in \mA_s$. Let
\begin{align}
\small
\bl p^* \in \argmax_{\bl p_{s_j}, j\in \{0,1,\dots w\}} 
\Bigg\{
&\sum_{a_i\in U}\bl p_{s_j}(a_i)\sum_{s'\in \mS}\mP(s'\mid s,\bl p_{ii})W(\pi',s') 
\Bigg\},
\end{align} 
breaking ties arbitrarily. Next, set $\pi'(s)= \bl p^*$. We have that
\begin{align*}
W(\pi,s)&=\sum_{s'\in \mS} \mP(s'\mid s,\pi)W(\pi,s') \\
&=\sum_{a_i\in U}\bl p_s(a_i)\sum_{s'\in \mS}\mP(s'\mid s,\bl p_{ii})W(\pi,s') \\
&\leq \argmax_{s_j\in \{s_0,s_1,\ldots ,s_w\}} 
\Bigg\{
\bigg(
\sum_{a_i\in U}\bl p_{s_j}(a_i)\sum_{s'\in \mS}\mP(s'\mid s,\bl p_{ii})W(\pi,s') 
\bigg)
\Bigg\} \\
&\leq \argmax_{s_j\in \{s_0,s_1,\ldots ,s_w\}} 
\Bigg\{
\bigg(
\sum_{a_i\in U}\bl p_{s_j}(a_i)\sum_{s'\in \mS}\mP(s'\mid s,\bl p_{ii})W(\pi',s') 
\bigg)
\Bigg\}\\
&= W(\pi',s).
\end{align*}
\end{proofof}

\begin{algorithm}[t]
\begin{algorithmic}[1]
\caption{Optimal Policy $\pi^*$ for the GMDP \label{alg:optimal policy}}
\REQUIRE{an instance $\langle \mS, \mA, \mP, \mR \rangle$.}
\ENSURE{an optimal policy $\pi^*$.}
\FOR{every non-terminal state $s=(O,U)\in\mS$}
\IF{$s=s_0$}
\STATE $\pi^*(s) \gets \bl p_{11}$.
\ELSE
\STATE $\pi^*(s) \gets \bl p_{i^*r^*}$ such that
{	
{\small
\begin{align}
\label{eq:i star and r star}
(i^*,r^*) \in \argmax_{\substack{(i,r)\in U \times U, \\ \bl p_{ir}\in \mA_s}}
\Bigg\{
\left(
1-\frac{\ind_{i=r}}{2}
\right)
\bigg(
&\bl p_{ir}(i)\sum_{s'\in \mS} \mP(s'|s,\bl p_{ii})W(\pi^*,s')+\bl p_{ir}(r)\sum_{s'\in \mS} \mP(s'|s,\bl p_{rr})W(\pi^*,s')
\bigg)
\Bigg\}.
\end{align}}%
}
\ENDIF
\ENDFOR
\end{algorithmic}
\end{algorithm}

\begin{proofof}{Theorem \ref{thm:optimal policy}} 
Fix an arbitrary policy $\pi$. We prove the claim by iterating over all states in an increasing order of the number of elements of $U$. We use induction to show that the constructed $\pi^*$ indeed satisfies the assertion. For convenience, we restate $\pi^*$ elaborately in Algorithm~\ref{alg:optimal policy}.

For every $s=(O,U)\in S$ such that $\abs{U}=1$, the claim holds trivially. To see this, recall that if $s$ is terminal, $\mA_s=\emptyset$; otherwise, the unique element in $\mA_s$ is the action that assigns probability 1 to the sole element in $U$. Either way, $W(\pi^*,s)= W(\pi,s)$.

Assume the assertion holds for every $\abs{U}\leq j$; namely, that $W(\pi^*,s)\geq W(\pi,s)$ for all $s= (O,U)\in \mS$ with $\abs{U}\leq j$. If $s$ is a terminal state, then we are done. Else, we shall make use of the following claim, which shows that every action in $\mA_s$ can be viewed as a weighted sum over the elements of $\{\bl p_{i,r} \in \mA_s\}$.
\begin{claim}
\label{claim:pi to succinct}
For any $s\in \mS$ and $\bl p\in \mA_s$, there exist coefficients $(z_{i,r})_{(a_i,a_r)\in U\times U}$ such that
\begin{itemize}
\item $z_{i,r} \geq 0$,
\item $\sum_{(a_i,a_r)\in U\times U}z_{i,r} =1$, and
\item $\bl p = \sum_{\bl p_{ir}\in \mA_s}z_{i,r}\bl p_{ir}$.
\end{itemize}
\end{claim}
The proof of the claim appears below this proof. In particular, Claim \ref{claim:pi to succinct} suggests that $\pi(s)$, which is valid and thus $\pi(s)\in \mA_s$ w.p. 1, can be presented as a weighted sum over all pairs $\bl p_{ir}\in \mA_s$. Finally,
\begin{align*}
W(\pi,s)&=\sum_{s'\in \mS} \mP(s'\mid s,\pi)W(\pi,s') \\
&=\sum_{a_i\in U}\pi(s)(a_i)\sum_{s'\in \mS}\mP(s'\mid s,\bl p_{ii})W(\pi,s') \\
&= \sum_{a_i\in U}\sum_{a_r\in U:\bl p_{ir}\in \mA_s}\left(
1-\frac{\ind_{i=r}}{2}
\right)(z_{i,r}\bl p_{ir}(i)+z_{r,i}\bl p_{ri}(i))\sum_{s'\in \mS}\mP(s'\mid s,\bl p_{ii})W(\pi,s') \\
&= \sum_{a_i\in U}\sum_{a_r\in U:\bl p_{ir}\in \mA_s}z_{i,r}\left(
1-\frac{\ind_{i=r}}{2}
\right)\left(\bl p_{ir}(i)\sum_{s'\in \mS}\mP(s'\mid s,\bl p_{ii})W(\pi,s')+ \bl p_{ir}(r)\sum_{s'\in \mS}\mP(s'\mid s,\bl p_{rr})W(\pi,s') \right)\\ 
&\leq \argmax_{{a_i,a_r\in U, \bl p_{ir}\in \mA_s}}\left\{
\left(
1-\frac{\ind_{i=r}}{2}
\right)
\left(
\bl p_{ir}(i)\sum_{s'\in \mS} \mP(s'|s,\bl p_{ii})W(\pi^*,s')
+\bl p_{ir}(r)\sum_{s'\in \mS} \mP(s'|s,\bl p_{rr})W(\pi^*,s')
\right)
\right\},\\ 
&= W(\pi^*,s),
\end{align*}
where the last equality follows since $\pi^*(s)=\bl p_{i^*,r^*}$ and by the definition of $(i^*,r^*)$ given in Equation (\ref{eq:i star and r star}). To sum, the constructed $\pi^*$ satisfies $W(\pi,s) \leq W(\pi^*,s)$ for every state $s$.
\end{proofof}

\begin{proofof}{Claim \ref{claim:pi to succinct}}
To ease readability, we shall use the notation $\alpha=\alpha(O)$ and  $d_i=\abs{\alpha-\mu_i}$ in this proof. Let $s$ be an arbitrary state and $\bl p\in \mA_s$ be an arbitrary action. Notice that $\bl p$ could be described as 
\begin{equation}
\label{eq in claim double star}
\bl p = \sum_{a_i\in U}v_i\cdot\bl p_{ii} +\sum_{\bl p_{ir}\in \mA_s}z_{i,r}\bl p_{ir},
\end{equation}
where $v_i= \bl p(i)$ and $z_{i,r}=0$ for every $a_i,a_r\in U$ such that $\bl p_{ir}\in \mA_s$. We now describe a procedure that shifts mass from the set $(v_i)_i$ to $(z_{i,r})_{i,r}$, while still satisfying the equality in Equation (\ref{eq in claim double star}). Each time we apply this procedure we decrease the value of one or more elements from $(v_i)_i$ and increase one or more elements from $(z_{i,r})_{i,r}$ by the same quantity. As a result, when it converges (assuming that it does), namely when $\sum_i v_i=0$, we are guaranteed that all the conditions of the claim hold. Importantly, throughout the course of this procedure, the following inequalities hold
\begin{equation}
\label{eq: procedure invariant 1}
\sum_{a_i\in U}v_i\cdot\mu_i \geq \alpha \sum_{a_i\in U}v_i.
\end{equation}
\begin{equation}
\label{eq: procedure invariant 2}
\sum_{a_i\in U}v_i+\sum_{\bl p_{ir}\in \mA_s}z_{i,r}=1.
\end{equation}
For the initial set of $(v_i)_i$ Equations (\ref{eq: procedure invariant 1})-(\ref{eq: procedure invariant 2}) trivially hold due to the way we initialize $(v_i)_i$ and since $\bl p\in \mA_s$ implies that
\begin{equation}
\label{eq: in claim, ir}
\sum_{a_i\in U}\bl p(i)\mu_i \geq \alpha.
\end{equation}
In each step of the procedure, we use the prime notation to denote the coefficients in the end of that step. The procedure operates as follows:
\begin{itemize}
\item If $v_i=0$ for every $a_i\in U$, the claim holds.
\item Else, if for every $i$ such that $v_i>0$, $\mu_i\geq \alpha$, then for every $i$ with $v_i>0$ set $z'_{i,i}=z_{i,i}+v_i$ and set $v'_i=0$. Notice that after this change Equations (\ref{eq in claim double star})--(\ref{eq: procedure invariant 2}) still hold.
\item There exists $i$ with $\mu_i<\alpha$ and $v_i>0$. Consequently, since Equation (\ref{eq: procedure invariant 1}) holds, there must exist $r$ such that $\mu_r >\alpha$ and $v_r>0$. We divide the analysis into three sub-cases, depending on the relation between $\frac{d_{r}}{d_{i}}$ and $\frac{v_i}{v_r}$.  
\begin{enumerate}
\item $\frac{d_{r}}{d_{i}} > \frac{v_i}{v_r}$: we replace $v_{i}$, $v_{r}$ and $z_{i,r}$   with   $v'_i$ , $v'_r$ and $z'_{i,r}$  such that $v'_{i} = 0$  , $v'_{r}= v_{r} -v_{i}\frac{d_{i}}{d_{r}}=v_r+v_i-\frac{v_i}{\bl p_{ir}(i)}$ and $z'_{i,r} = z_{i,r} + v_i\frac{d_{i}+d_{r}}{d_{r}}= z_{i,r} +\frac{v_i}{\bl p_{ir}(i)}$. Clearly, after this modification the new coefficients are non-negative. To show that Equation (\ref{eq in claim double star}) still holds, we need to show that $\bl p(i), \bl p(r)$ can be decomposed using the new coefficients. Notice that
\begin{align*}
\bl p(i)&=v_i+\sum_{j:\bl p_{i,j}\in \mA_s}z_{i,j}\bl p_{ij}(i)+\sum_{j:\bl p_{j,i}\in \mA_s}z_{j,i}\bl p_{ji}(i)\\
&=v_i'+v_i+z_{i,r}\bl p_{ir}(i)+\sum_{j:j\neq r,\bl p_{i,j}\in \mA_s}z_{i,j}\bl p_{ij}(i)+\sum_{j:\bl p_{j,i}\in \mA_s}z_{j,i}\bl p_{ji}(i)\\
&=v_i'+v_i\frac{\bl p_{ir}(i)}{\bl p_{ir}(i)} +z_{i,r}\bl p_{ir}(i)+\sum_{j:j\neq r,\bl p_{i,j}\in \mA_s}z'_{i,j}\bl p_{ij}(i)+\sum_{j:\bl p_{j,i}\in \mA_s}z'_{j,i}\bl p_{ji}(i)\\
&=v_i'+\left(\frac{v_i}{\bl p_{ir}(i)} +z_{i,r}\right)\bl p_{ir}(i)+\sum_{j:j\neq r,\bl p_{i,j}\in \mA_s}z'_{i,j}\bl p_{ij}(i)+\sum_{j:\bl p_{j,i}\in \mA_s}z'_{j,i}\bl p_{ji}(i)\\
&=v_i'+z'_{i,r}\bl p_{ir}(i)+\sum_{j:j\neq r,\bl p_{i,j}\in \mA_s}z'_{i,j}\bl p_{ij}(i)+\sum_{j:\bl p_{j,i}\in \mA_s}z'_{j,i}\bl p_{ji}(i)\\
&=v_i'+\sum_{j:\bl p_{i,j}\in \mA_s}z'_{i,j}\bl p_{ij}(i)+\sum_{j:\bl p_{j,i}\in \mA_s}z'_{j,i}\bl p_{ji}(i).
\end{align*}
Similarly,
\begin{align*}
\bl p(r)&=v_r+\sum_{j:\bl p_{r,j}\in \mA_s}z_{r,j}\bl p_{rj}(r)+\sum_{j:\bl p_{j,r}\in \mA_s}z_{j,r}\bl p_{jr}(r)\\
&=v'_r-v_i+\frac{v_i}{\bl p_{ir}(i)}+z_{i,r}\bl p_{ir}(r)+\sum_{j:\bl p_{r,j}\in \mA_s}z_{r,j}\bl p_{rj}(r)+\sum_{j:j\neq i,\bl p_{j,r}\in \mA_s}z_{j,r}\bl p_{jr}(r)\\
&=v'_r+\frac{v_i(1-\bl p_{ir}(i))}{\bl p_{ir}(i)} +z_{i,r}\bl p_{ir}(r)+\sum_{j:\bl p_{r,j}\in \mA_s}z'_{r,j}\bl p_{rj}(r)+\sum_{j:j\neq i,\bl p_{j,r}\in \mA_s}z'_{j,r}\bl p_{jr}(r)\\
&=v'_r+\frac{v_i\bl p_{ir}(r)}{\bl p_{ir}(i)} +z_{i,r}\bl p_{ir}(r)+\sum_{j:\bl p_{r,j}\in \mA_s}z'_{r,j}\bl p_{rj}(r)+\sum_{j:j\neq i,\bl p_{j,r}\in \mA_s}z'_{j,r}\bl p_{jr}(r)\\
&=v'_r+\left(\frac{v_i}{\bl p_{ir}(i)} +z_{i,r}\right)\bl p_{ir}(r)+\sum_{j:\bl p_{r,j}\in \mA_s}z'_{r,j}\bl p_{rj}(r)+\sum_{j:j\neq i,\bl p_{j,r}\in \mA_s}z'_{j,r}\bl p_{jr}(r)\\
&=v'_r+\sum_{j:\bl p_{r,j}\in \mA_s}z'_{r,j}\bl p_{rj}(r)+\sum_{j:\bl p_{j,r}\in \mA_s}z'_{j,r}\bl p_{jr}(r).
\end{align*}

As a result, Equation (\ref{eq in claim double star}) holds. As for Equation (\ref{eq: procedure invariant 1}), observe that
\begin{align*}
\sum_{a_j\in U}v'_j\cdot\mu_j &= v_i'\mu_i+v_r'\mu_r+ \sum_{j\notin \{i,r\} ,a_j\in U}v'_j\cdot\mu_j\\
&= v_r\mu_r-v_i\mu_r\frac{d_i}{d_r} +\sum_{j\notin \{i,r\} ,a_j\in U}v_j\cdot\mu_j \\
&= v_r\mu_r+v_i\mu_i-v_i \mu_i-v_i\mu_r\frac{d_i}{d_r} +\sum_{j\notin \{i,r\} ,a_j\in U}v_j\cdot\mu_j \\
&= -v_i \mu_i-v_i\mu_r\frac{d_i}{d_r} +\sum_{a_j\in U}v_j\cdot\mu_j \\
&\geq -v_i \mu_i-v_i\mu_r\frac{d_i}{d_r} +\alpha\sum_{a_j\in U}v_j \\
&= -v_i \mu_i-v_i\mu_r\frac{d_i}{d_r}+\alpha\cdot v_i+\alpha\cdot v_r +\alpha\sum_{j\notin \{i,r\} ,a_j\in U}v_j \\
&= -v_i \mu_i-v_i\mu_r\frac{d_i}{d_r}+\alpha\cdot v_i+\left(\alpha\cdot v_r'+\alpha\cdot v_i\frac{d_i}{d_r} \right)+\alpha\cdot v_i'+\alpha\sum_{j\notin \{i,r\} ,a_j\in U}v_j \\
&= -v_i \mu_i-v_i\mu_r\frac{d_i}{d_r}+\alpha\cdot v_i+\alpha\cdot v_i\frac{d_i}{d_r} +\alpha\sum_{a_j\in U}v_j' \\
&= v_i\left( -\mu_i-\mu_r\frac{d_i}{d_r}+\alpha+\alpha\frac{d_i}{d_r} \right)+\alpha\sum_{a_j\in U}v_j' \\
&= v_i\left(d_i-d_r\frac{d_i}{d_r}\right)+\alpha\sum_{a_j\in U}v_j' \\
&=\alpha\sum_{a_j\in U}v_j';
\end{align*}

hence, Equation (\ref{eq: procedure invariant 1}) holds. Finally, $v_i+v_r+z_{i,r}=v'_i+v'_r+z'_{i,r}$ while all other coefficients are left unchanged; thus Equation (\ref{eq: procedure invariant 2}) holds as well.
\omer{I put the other cases in a separate file.}
\item $\frac{d_{r}}{d_{i}} < \frac{v_i}{v_r}$: the analysis is similar to the previous case and hence omitted.
\item $\frac{d_{r}}{d_{i}} = \frac{v_i}{v_r}$: 
the analysis is similar to the first case and hence omitted.
\end{enumerate}
\end{itemize}
This concludes the proof.
\end{proofof}

\section{Omitted Proofs from Subsection \ref{sec:algorithm}}

\begin{proofof}{Proposition \ref{prop:is EAIR}}
We need to show that Inequality (\ref{eq: ir condition}) holds for every history $h$. Since \mainalg operates in phases, it would be convenient to divide the arguments into these three phases, according to which phase $h$ belongs.
\begin{itemize}
\item Exploration phase: the recommendation is based on the action of $\pi^*$, the optimal policy of the GMDP in Subsection \ref{subsec:explore}. If $h$ is the empty history, then it is translated to $s_0$, and $\pi^*$ selects $a_1$ w.p. 1. Otherwise, due to  Equation (\ref{eq:gmdp EAIR}) the action space of the GMDP is restricted to distributions over the unobserved arms with expectation greater or equal to the observed value $R_1$.  As a result, in both cases Inequality (\ref{eq: ir condition}) holds.
\item Experience phase: in this phase, \mainalg$(h)$ is a distribution over two arms, $\tilde r$ and $i$, with $R_{\tilde r}$ greater than the obtained value $R_1$ of arm $a_1$. Further, $X_i > R_{\tilde r}$ with positive probability, or otherwise arm $a_i$ would have been discarded (Lines \ref{alg: dominated}--\ref{alg: while explore dominated continue}). If, in addition, $\mu_i \geq R_1$, then the If sentence in Line \ref{alg:while if conditions} would select arm $a_i$ with probability 1, satisfying Inequality (\ref{eq: ir condition}). On the other hand, if $\mu_i < R_1$, then \mainalg selects arm $a_i$ w.p. $\frac{R_{\tilde r}-R_1}{R_{\tilde r}-\mu_i}$, and $a_{\tilde r}$ with the remaining probability (Lines \ref{alg:draw}--\ref{alg: best arm}); hence expected value of \mainalg$(h)$ is
\[
\mu_i \cdot \frac{R_{\tilde r}-R_1}{R_{\tilde r}-\mu_i}+R_{\tilde r}\left(1- \frac{R_{\tilde r}-R_1}{R_{\tilde r}-\mu_i}\right),
\]
which is greater or equal to $R_1$.
\item Exploit phase: in this phase \mainalg$(h)$ is a deterministic selection of one arm --- the most rewarding one. Since the value of arm $a_1$, $R_1$ was observed before (as mentioned for the exploration phase), the arm $a_{i^*}$ selected in Line \ref{alg: best arm}, satisfies $R_{i^*} >R_i$.
\end{itemize}
\end{proofof}

\subsection{Optimality}
\begin{proofof}{Theorem \ref{theorem:ic-ir upper}}
To facilitate the proof, we introduce the following definitions: given a mechanism $M$ and a history $h$, we say that $M$ is \textit{fruitless} w.r.t. $h$ if $M(h)$ gives a positive probability to at least one  observed arm $a_i$, $i\neq 1$, with $R_i\leq R_1$, i.e., reward that is at most $R_1$ (notice that it implies that $a_1$ and $a_i$ were observed). In addition, we say that a history $h$ is \textit{auspicious} if an action with reward greater than that of $a_1$ is observed under $h$.

We are ready to begin the proof. Let $M$ be an arbitrary mechanism, and for the sake of the proof fix the number of agents, and only consider histories of length of at most $n$. The proof contains three steps. In Step 1 we slightly modify $M$, resulting in a new mechanism $\mm 1$ that attains a social welfare at least as high as that of $M$, and is still EAIR. In Step 2, we modify $\mm 1$ to use an oracle whenever it reaches an auspicious history. As we show, the resulting mechanism, $\mm 2$ has an improved social welfare, $SW(\mm 2)\geq SW(\mm 1)$. Finally, in Step 3 we show that the social welfare of $\mm 2$ is at most $W(\pi^*,s_0)$. 

\paragraph{Step 1:}  In this step we construct a modification of $M$ with at least the same social welfare, which is not fruitless on any history $h$. We define a mechanism $\mm 1$ that receives $M$ as a black box and uses it for recommendations. $\mm 1 $ is defined as follows:
\begin{enumerate}
\item Let $\tih$ be the empty history. Act as $M(\tih)$ and update $\tih$ accordingly.
\item While the length of $\tih$ is less than $n$:
\begin{enumerate}[label=\text{2.\arabic*},ref=\text{2.\arabic*}]
	\item Draw $a_i\sim M(\tih)$. If the reward of $a_i$ was already observed and $R_i \leq R_1$, recommend $a_1$ and set $\tih = \tih \oplus (a_i,R_i)$. Else, act as $M(\tih)$ and update $\tih$ accordingly.
\end{enumerate}
\end{enumerate}
It is straightforward to see that $\mm 1$ satisfies the EAIR condition, and that $SW(\mm 1) \geq SW(M)$. 
\paragraph{Step 2:} In this step, we present a non-feasible mediator $\mm 2$ that modifies the way $\mm 1$ operates on auspicious histories. $\mm 2$ uses an oracle that hints the best arm.

More concretely, $\mm 2$ is defined as follows:
\begin{enumerate}
\item Let $\tih$ be the empty history. Act as $\mm 1(\tih)$ and update $\tih$ accordingly.
\item While $\tih$ is not auspicious:
\begin{enumerate}[label=\text{2.\arabic*},ref=\text{2.\arabic*}]
\item Act as $\mm 1(\tih)$ and update $\tih$ accordingly.
\end{enumerate}
\item If $\tih$ is auspicious:
\begin{enumerate}[label=\text{3.\arabic*},ref=\text{3.\arabic*}]
\item \label{mm 4 oracle hint} Use an oracle to reveal the best arm, $a^*$. From here on, recommend $a^*$ to all users.
\end{enumerate}
\end{enumerate}

Notice that $\mm 2$ is EAIR for every non-auspicious history, but not EAIR in general; for this reason, it is not feasible. Moreover, it holds that $SW(\mm 2) \geq SW(\mm 1)$.

\paragraph{Step 3:} 
The final step is to claim that the resulting mechanism $\mm 2$ cannot get more than the optimal value of the GMDP in Section \ref{sec:aux mdp}. However, the GMDP does not allow selecting $a_1$, so we have to have some minor modifications.

This step is structured as follows. First, formally define a modified version of the GMDP presented in Section \ref{sec:aux mdp}, with minor modifications. We call the new GMDP \textit{Repeated GMDP}, or R-GMDP for abbreviation to distinguish between the two. Then, we show that the best achievable value in the R-GMDP is exactly $W(\pi^*,s_0)$. The final step is mapping $\mm 2$ obtained in Step 2 to a non-stationary strategy in the R-GMDP, which achieves at least as as the social welfare of $\mm 2$, that is $SW(\mm 2)$. The claim then follows since the policy constructed using $\mm 2$ cannot obtain more that $W(\pi^*,s_0)$.

Consider the following R-GMDP: \footnote{The crucial difference between R-GMDP and GMDP is in the action space and the transition probabilities, colored in red for readability.}
\begin{itemize}[leftmargin=*]
\item $\mS$ is a finite set of states. Each state $s$ is a pair $(O,U)$, where $O \subseteq \{(a,c)\mid a\in A, c\in H \}$ is the set of arm--reward pairs that have been observed so far.  $U\subseteq A$ is the set of arms not yet explored. The initial state is thus $s_0=(\emptyset, A)$. For every non-empty set of pairs $O$ we define $\alpha(O)$ to be the reward observed for arm $a_1$ (that can be obtained several times, as we explain shortly), and $\beta(O)= \max_{c:\exists a, (a,c)\in O}c$ to be the maximal reward observed.
\item $\mA=\bigcup_{s \in \mS} \mA_s$ is an infinite set of actions. For each $s=(O,U)\in \mS$, $\mA_s$ is defined as follows:
	\begin{enumerate}
	\item If $s=s_0$, then $\mA_{s_0}=\Delta(\{a_1\})$: i.e., a deterministic selection of $a_1$.
	\item Else, if $\alpha(O) < \beta(O) $, then $\mA_s=\emptyset$. 
	\item Otherwise, $\mA_s$ is a subset of \red{ $\Delta(U\cup \{a_1\})$}, such that $\bl p \in \mA_s$ if and only if 
	\[
    \sum_{a_i\in \red{U\cup \{a_1\}}} \bl p(a_i) \mu_{a_i} \geq \alpha(O).
    \]
\end{enumerate}
We denote by $\mS_T$ the set of \textit{terminal} states, namely $\mS_T =\{s\in\mS \mid \mA_s =\emptyset \}$.

\item $\mP$ is the transition probability function. Let $s=(O,U) \in \mS$, and let $s'=(O',U')$ such that $O' =O\cup \{(a_i,c)\}$ and $U'=U\setminus\{a_i\}$ for some $a_i\in \red{U \cup \{a_1\}}, c\in [H]^+$.

Then, the transition probability from $s$ to $s'$ given an action $\bl p$ is defined by

\[
\mP(s'|s,\bl p)= 
\begin{cases}
\bl p (a_i) \Pr(X_i=c) & a_i \in U \\
\red{\ind_{c=\alpha(O)}}                      & \red{a_i=a_1}
\end{cases}.
\]
If $s'$ is some other state that does not meet the conditions above, then let $\mP(s'|s,\bl p)=0$ for every $\bl p \in \mA_s$.
\item $\mR : \mS_T \rightarrow \mathbb{R}$ is the reward function, defined on terminal states only. For each terminal state $s=(O,U) \in \mS_T$,
\[
\mR(s)=
\begin{cases}
\alpha(O) &  \alpha(O)=\beta(O)\\
\E\left[\max\left\{\beta(O), \max_{a_{i'}\in U}X_{i'}) \right\} \right]  & \alpha(O)<\beta(O).
\end{cases}.
\]
\end{itemize}

Next, we prove that there exists an optimal policy for the R-GMDP with a significantly reduced support.
\begin{lemma}\label{lemma-r-gmdp}
For every policy $\pi$ for R-GMDP, there exists a stationary policy $\pi'$ such that 
\begin{enumerate}
    \item $\pi'(s)=\pi'(s')$ for every pair of states $s=(O,U)$ and $s'=(O',U)$ with $\alpha(O)=\alpha(O')$ and  $\beta(O)=\beta(O')$.
    \item For every state s,  $W(\pi',s)\geq W(\pi,s)$.
\end{enumerate}
\end{lemma}
The proof of the lemma is identical to the proof of Lemma \ref{lemma:gmdp reduction} and hence omitted. Lemma \ref{lemma-r-gmdp} suggests that we can focus on strategies that distinguish between states based on $U,\alpha(O)$ and $\beta(O)$ solely. The reduced state space does allows self loop  by selecting $a_1$, without having any effect on the reward. It is thus straightforward to see that an optimal strategy that ignores $a_1$ exists, with a reward of exactly $W(\pi^*,s_0)$.

Notice that $\mm 2$ defines a non-stationary policy $\pi$ for the R-GMDP, by mimicking the actions (distributions) $\pi$ selects. When $\mm 2$ gets to an auspicious history or could not explore anymore, the policy $\pi$ gets to a terminal state and obtains a reward. Each time $\mm 2$ directs an agent, that agent gets at most the maximal reward $\mm 2$ discovered; hence, $SW(\mm 2)$ is less or equal to the reward obtained by that non-stationary policy $\pi$, which is at most $W(\pi^*,s_0)$.

This completes the proof of the theorem.
\end{proofof}

\begin{proofof}{Lemma \ref{lemma:our alg sw is best}}
Let $N_1, N_2$ denote the r.v. representing the number of agents in the explore and experience phases, respectively. Notice that the definition of social welfare given in Equation \ref{eq: social welfare} can be interpreted as
\begin{align}\label{eq: social with three}
SW(\textnormal{\algname}) = \frac{1}{n}\left( \E\left(\sum_{l=1}^{N_1+N_2} X_{M(h_l)} \right) +\E\left(\sum_{l''=N_1+N_2+1}^n X_{M(h_{l''})} \right) \right).
\end{align}
Observe that every agent in the explore and experience phases obtains the reward of arm $a_1$ in expectation. Moreover, every agent in the exploit phase obtains $W(\pi^*,s_0)$ in expectation; hence, Equation (\ref{eq: social with three}) can be rearranged as
\begin{align*}\label{eq: social with three after one}
SW(\textnormal{\algname}) &= \frac{1}{n}\left( \E\left(\sum_{l'=1}^{N_1+N_2} X_1 \right) +\E\left(\sum_{l''=N_1+N_2+1}^n W(\pi^*,s_0) \right) \right) \\
&= \frac{1}{n}\left( \mu_1\E\left(N_1+N_2\right) +W(\pi^*,s_0)\E\left(n-N_1-N_2 \right) \right)\\
&=W(\pi^*,s_0)- \frac{1}{n}\E\left(N_1+N_2\right)(W(\pi^*,s_0)-\mu_1).
\end{align*}
To finalize the proof, recall that $N_1\leq K$ almost surely since there are $K$ arms that could be explored, and on every step in the exploration phase exactly one arm gets explored. Moreover, due to Observation \ref{prop:dominated} it holds that $\E(N_2)\leq KH$; hence,
\begin{align*}
SW(\textnormal{\algname}) & \geq  W(\pi^*,s_0)- \frac{1}{n} \left(K+KH\right)(W(\pi^*,s_0)-\mu_1) .
\end{align*}
\end{proofof}
 
\section{Omitted Proofs from Section \ref{sec:ic}}
\begin{proofof}{Theorem \ref{theorem: ic fee}}
It is immediate to see that \mainicalg is EAIR.  Satisfying EAIR follows from mixing \mainalg, which is EAIR, with \gre, which satisfies the delegate property and hence also the EAIR constraint, and recommendations of $a_1$.

Moreover, \mainicalg is asymptotically optimal since, after finitely many agents, its recommendations will coincide with those of \mainalg, and \mainalg is asymptotically optimal. While \mainicalg is not exploiting (Line \ref{alg2:if exploits}), its recommendations coincide with those of \mainalg at least once per phase. Since the expected exploration time of \mainalg is $O\left(\frac{KH^2}{n}  \right)$ (see Lemma \ref{lemma:our alg sw is best}), \mainicalg explores for $O\left(\frac{BKH^2}{n}  \right)$ rounds in expectation. 
 
Showing that \mainicalg satisfies IC is trickier. We divide the analysis to several parts:

\begin{itemize}
\item The first agent gets $a_1$, which is the a-priori best action.
\item Agents $2,\dots,K$ either get recommendations from \gre (Line \ref{alg2: r1 is bad}) or are recommended $a_1$ (Line \ref{alg2: r1 after one}). In the former, agents get the best arm known to the mechanism. In the latter, the only new information agents could learn is that $X_1\geq \mu_K$; thus, for every $a_j\neq a_1$ it holds that
\[
\E(X_1-X_j\mid X_1\geq \mu_K)\stackrel{iid}{\geq} \E(X_1-X_j)\geq 0.
\]
Agents cannot know if they are being recommended by Line \ref{alg2: r1 is bad} or Line \ref{alg2: r1 after one}, but in both cases they are better off with accepting the recommendation; hence, IC holds for agents $2,\dots,K$.
\item Agents $K+1,\dots,n$, and the recommended arm is $a_1$. This case might be trivial at first glance, but it is not as innocent. \mainicalg can recommend $a_1$ via Lines \ref{alg2:exploits} and \ref{alg2: recommend a on in phase}. In both cases, we know that $a_1$ is the best among all the explored arms. Nevertheless, there could still be unexplored arms with an expected value greater than $R_1$. One such scenario is when $R_1$ revealed by the first agent yielded $\mu_2 \leq R_1 \leq \mu_3$. In this case, \mainicalg recommends 
agent $K+1$, assuming that she was not selected to be the exploring agent, arm $a_1$. Nevertheless, according to \mainicalg's information at that point, arm $a_2$ is the best arm. Recommending $a_2$ greedily might disallow the mechanism to explore more arms using the mixture \mainalg employs, which leads to sub-optimal social welfare.

Nevertheless, we will show that IC holds in this case as well. Fix an agent $l$ and some phase $k$, and assume \mainicalg recommended agent $l$ arm $a_1$. Let $E^l_{O}$ denote the event indicating that $O\subseteq A$ arms where observed just before agent $l$ arrives, and $X_1 \geq X_i$ for every $a_i \in O$. Clearly, agent $l$ does not know whether $E^l_{O}$ occurs or not, but she can  compute the occurrence probability. We have that
{\thinmuskip=0mu
\medmuskip=0mu plus 0mu minus 0mu
\thickmuskip=0mu plus 0mu
\begin{align}\label{eq: for rec of one}
&\E(X_1-X_j\mid M=a_1)=\E(X_1-X_j\mid X_1 > \mu_2)\Pr(X_1>\mu_2)+\sum_{O\subseteq A}\E(X_1-X_j\mid X_1 \leq \mu_2,E^l_O)\Pr(X_1\leq\mu_2,E^l_O).
\end{align}}%
In addition,
\[
\E(X_1-X_j\mid X_1 \leq \mu_2,E^l_O) \geq \E(X_1-X_j\mid X_1 \leq \mu_2).
\]
This inequality follows immediately if $a_j\in O$. Otherwise, if $a_j\notin O$, due to the i.i.d. assumption,$X_1-X_j$ could only increase conditioning on $E^l_O$; hence,
{\thinmuskip=0mu
\medmuskip=0mu plus 0mu minus 0mu
\thickmuskip=0mu plus 0mu
\begin{align*}
\textnormal{Eq.}(\ref{eq: for rec of one})&\geq \E(X_1-X_j\mid X_1 > \mu_2)\Pr(X_1>\mu_2)+\E(X_1-X_j\mid X_1 \leq \mu_2)\Pr(X_1\leq\mu_2).\nonumber\\
&=\E(X_1-X_j\mid M=a_1) \geq 0.
\end{align*}}%
We conclude that agents $K+1,\dots,n$ follow \mainicalg when it recommends $a_1$.
\item Agents $K+1,\dots,n$, and the recommended arm is $a_i\neq a_1$. Fix an agent $l$ and some phase $k$, and assume \mainicalg recommended agent $l$ arm $a_i\neq a_1$. We need to show that for every $a_j$, it holds that $\E\left[X_i-X_j\mid M=a_i\right] \geq 0$. Due to Assumption \ref{assumption ic}, there exists $\xi>0,\gamma>0$ such that
\[
\forall i\in [K]:\qquad \Pr(\forall i'\in [K]\setminus \{i\}:\mu_i -X_{i'} > \xi  )>\gamma.
\]
In words, Assumption \ref{assumption ic} guarantees that with positive probability $\gamma$, all arms but $i$ have a reward that is less than $\mu_i$ by at least $\xi$. Denote this event by $\ind_{a_i}$. If $\ind_{a_i}$ occurs, we are guaranteed that arm $a_i$ will be explored in Line \ref{alg2: r1 is bad}. Moreover, denote by $\ind_{l,exp}$ the event that agent $l$ is the agent selected by \mainicalg to explore in Line \ref{alg2: pick to explore}. We have that
{\thinmuskip=0mu
\medmuskip=0mu plus 0mu minus 0mu
\thickmuskip=0mu plus 0mu
\begin{align}\label{eq: ic for B}
&\E\left[X_i-X_j\mid M=a_i \right]=\E\left[X_i-X_j\mid M=a_i, \ind_{l,exp}\right]\Pr(\ind_{l,exp})+\E\left[X_i-X_j\mid M=a_i,\overline{\ind_{l,exp}}  \right]\Pr(\overline{\ind_{l,exp}})\nonumber\\
&\geq \frac{-H}{B}+\E\left[X_i-X_j\mid M=a_i,\overline{\ind_{l,exp}} ,\ind_{a_i}  \right]\Pr(\overline{\ind_{l,exp}},\ind_{a_i} )+\underbrace{\E\left[X_i-X_j\mid M=a_i,\overline{\ind_{l,exp}} ,\overline{\ind_{a_i}}  \right]}_{\geq 0}\Pr(\overline{\ind_{l,exp}},\overline{\ind_{a_i}} )\nonumber\\
&\geq \frac{-H}{B}+\frac{\xi \gamma (B-1)}{B},
\end{align}%
}%
and the latter is non-negative if $B\geq \frac{H}{\xi \gamma}+1$. 
\end{itemize}
Overall, we showed that every agent is better off by accepting \mainicalg's recommendation; hence, \mainicalg is IC.
\end{proofof}

\section{Omitted Proofs and Claims from Section \ref{sec:analysis}}

\subsection{Ex-Post Individual Rationality}\label{sec:EPIR}
Notice that EAIR mechanisms guarantee each agent the value of the default arm, but only in expectation. We now propose a more strict form of individual rationality, \textit{ex-post} individual rationality (EPIR).
\begin{definition}[\textit{Ex-Post} Individual Rationality]
A mechanism $M$ is \textit{ex-post}  individually rational (EPIR) if for every agent $l\in\{1, \ldots, n\}$, every value $R_1$ in the support of $X_1$, every history $h=(h_1,\ldots,h_{l-1}) \in \left(A\times \R_+ \right)^{l-1}$, and every arm $a_r$ such that $\Pr(M(h)=r)>0$, it holds that $
\E(X_{r} \mid h)  \geq  R_1.$
\end{definition}
Satisfying EPIR means that the mechanism never recommends an arm that is \textit{a priori} inferior to arm $a_1$. Noticeably, every EPIR mechanism is also EAIR, yet EPIR mechanisms are quite conservative, since they can only explore arms that yield expected rewards of at least the value $R_1$ obtained for $a_1$. 

An optimal EPIR mechanism is immediate in case of non-strategic agents; we denote by $\mep$ this intuitive mechanism. First, explore arm $a_1$, and observe $R_1$. Then, remove all arms $a_r$ with $\mu_r < R_1$, and name the obtained set $A'$. Then, proceed with \fulexp~ on $A' \cup \{a_1\}$. 

For the case of strategic agents, $\mep$ is not enough: agents might be reluctant to explore arms with a-priori low rewards. We propose \mainicepalg, which is an asymptotically optimal IC and EPIR mechanism. \mainicepalg relies on the same technique we use in Section \ref{sec:ic} and is outline in Algorithm \ref{mainicepalg}. 
\begin{theorem}\label{theorem: ic ep fee}
Let the phase length be $B= \ceil*{\frac{H}{\xi \gamma}}+1$. Under Assumption \ref{assumption ic}, $\textnormal{\mainicepalg}$ satisfies EPIR and IC. In addition,
$
SW_{n}(\textnormal{\mainicalg}) \geq \OPTEP-O\left(\frac{KH^3}{n\xi \gamma}  \right).
$
\end{theorem}
The proof of Theorem \ref{theorem: ic ep fee} is similar to that of Theorem \ref{theorem: ic fee}, and is hence omitted.
\begin{algorithm}[t]
\caption{IC EPIR Explore \& Exploit \label{mainicepalg}(IC-EP-FEE)}
\begin{algorithmic}[1]
\STATE Initialize an instance of $\mep$ and update it after every recommendation.\label{alg3:initialize} 
\STATE Recommend as \gre to agents $1,2,\dots, K$. \label{alg3: gre}
\STATE Split the remaining rounds into consecutive phases of $B$ rounds each. \label{alg3:split}
\FOR {phase $k=1,\dots$ }
\IF{$\mep$ exploits\label{alg3:if exploits}}
 {
\STATE follow $\mep$ \label{alg3:exploits} 
}
\ELSE {\label{alg3:else block explore}
\STATE Pick an agent $l(k)$ from the $B$ agents in this phase uniformly at random.
\STATE Every agent in this phase is recommended as \gre, except agent $l(k)$ who is recommended  according to $\mep$. \label{alg3: explore or exploit}
}
\ENDIF
\ENDFOR
\end{algorithmic}
\end{algorithm}

\subsection{Omitted Proofs from Subsection~\ref{subsec:welfare analysis}}

\begin{proofof}{Proposition \ref{prop:bound on EAIR}}
Let $X_1$ be such that $\Pr(X_1=1)=1$, and for every $i$ such that $2\leq i\leq K$ let
\[
X_i = \begin{cases}
0 & \text{ w.p. } 1-\frac{1}{H} +\epsilon \\
H & \text{ w.p. } \frac{1}{H} -\epsilon
\end{cases},
\]
for a small positive constant $\epsilon$. Clearly, $\mu_1=1$ while $\mu_i<1$ for $2\leq i \leq K$; hence, $\OPTEA=1$. On the other hand,
\begin{align*}
\OPT &= \E(\max_{1\leq i \leq K} X_i) = \Pr(\max_{2\leq i \leq K} X_i =H)H+\Pr(\max_{2\leq i \leq K} X_i =0)\cdot 1 \\
&= \left(1-(1-\frac{1}{H} +\epsilon)^{K-1}\right) H +(1-\frac{1}{H} +\epsilon)^{K-1}.
\end{align*}
Taking $\epsilon$ to zero, we get that $\OPT$ is arbitrarily close to
\begin{align*}
& \left(1-(1-\frac{1}{H})^{K-1}\right) H +(1-\frac{1}{H} )^{K-1} = H\left(1-(1-\frac{1}{H})^K\right).
\end{align*}
Finally, we use the fact that $e^{-x} \geq (1-\frac{x}{n})^n$ whenever $\abs{x} \leq n$. By setting $n=K$ and $x=\frac{K}{H}$, we conclude that $e^{-\frac{K}{H}} \geq (1-\frac{1}{H})^K$; therefore,
\[
\frac{\OPT}{\OPTEA} \geq H\left(1-e^{-\frac{K}{H}} \right).
\]
\end{proofof}

\begin{proofof}{Proposition \ref{prop:bound on EAIR-EPIR}}
Let $X_1,X_2,\dots X_K$ such that
\[
X_1 = \begin{cases}
1 & \text{ w.p. } 1-\frac{1}{H-1}-\epsilon \\
H & \text{ w.p. } \frac{1}{H-1} +\epsilon
\end{cases}, \quad
X_2 =
\begin{cases}
2 & \text{ w.p. } 1
\end{cases}, \quad
\forall 3\leq i \leq K :\quad 
X_i = \begin{cases}
0 & \text{ w.p. } 1-\frac{1}{H} +\epsilon \\
H & \text{ w.p. } \frac{1}{H} -\epsilon
\end{cases}.
\]
It holds that
\begin{align*}
\OPTEP =  H\left(\frac{1}{H-1} +\epsilon  \right) + 2\left(1-\frac{1}{H-1}-\epsilon\right). 
\end{align*}
On the other hand,
\begin{align*}
\OPTEA =  H\left(\frac{1}{H-1} +\epsilon  \right)+2\left( 1-\frac{1}{H} +\epsilon  \right)^{K-2}+H\left(1-\left( 1-\frac{1}{H} +\epsilon  \right)^{K-2}  \right).
\end{align*}
Taking $\epsilon$ to zero, we get
\begin{align*}
\frac{\OPTEA}{\OPTEP} &=  \frac{ H\left(\frac{1}{H-1}\right)+2\left( 1-\frac{1}{H}  \right)^{K-2}+H\left(1-\left( 1-\frac{1}{H}   \right)^{K-2}  \right) }{3-\frac{1}{H-1}}  \\ 
& \geq \frac{(H+2)\left(1-\left( 1-\frac{1}{H}   \right)^{K-2}  \right)}{3} \\
& \geq \frac{H+2}{3}\left(1-e^{-\frac{K-2}{H}} \right).
\end{align*}
\end{proofof}

\begin{proofof}{Proposition \ref{prop:bound on EPIR-del}}
Let $X_1,X_2,\dots X_K$ such that
\[
X_1 = \begin{cases}
0 & \text{ w.p. } \frac{1}{2}-\epsilon \\
2 & \text{ w.p. } \frac{1}{2} +\epsilon
\end{cases}, \quad
X_2 =
\begin{cases}
1 & \text{ w.p. } 1
\end{cases}, \quad
\forall 3\leq i \leq K :\quad 
X_i = \begin{cases}
0 & \text{ w.p. } 1-\frac{1}{H} +\epsilon \\
H & \text{ w.p. } \frac{1}{H} -\epsilon
\end{cases}.
\]
For $\epsilon\rightarrow 0$. It holds that $
\OPTDEL=\frac{1}{2}\cdot 2+\frac{1}{2}\cdot 1=1.5$. On the other hand,
\begin{align*}
\OPTEP&=\frac{1}{2}\cdot 2+\frac{1}{2}\cdot \left(1(1-\frac{1}{H})^{K-2}+H\cdot (1-(1-\frac{1}{H})^{K-2}) \right)\\
&\leq 1+\frac{H}{2}\left( 1-e^{-\frac{K-2}{H}} \right);
\end{align*}
thus, $\frac{\OPTEP}{\OPTDEL} \geq \frac{H}{3}\left( 1-e^{-\frac{K-2}{H}}\right)$.
\end{proofof}

\begin{proposition} \label{prop:bound on EAIR uniform}
Fix $K,H\in \mathbb N$. Let $X_i\sim Uni[H]^+$, and let
$
X_1 =\begin{cases} Uni[H]^+ & \textnormal{w.p. } 1-\epsilon \\ H & \textnormal{w.p. } \epsilon \end{cases}
$
for arbitrarily small $\epsilon>0$. It holds that $\frac{\OPT}{\OPTEA} \leq \frac{8}{7}+O(\epsilon)$.
\end{proposition}

\begin{proofof}{Proposition \ref{prop:bound on EAIR uniform}} Assume for simplicity that $H$ is even. First, by simple probability tricks one can show that 
\[
\OPT = \E( \max_{1\leq i \leq K} X_i) = (1-\epsilon)\frac{K}{K+1}H+\epsilon H=\frac{K}{K+1}H+O(\epsilon).
\]
Second, since $\E(X_1)>\E(X_i)$ for every $i\in \{2,\dots K\}$, any EAIR mechanism must explore $X_1$ first. Notice that $\max_{2\leq i \leq K} \mu _i = \frac{H}{2}$; thus,
\begin{align*}
\OPTEA &= \Pr(X_1 > \frac{H}{2})\E(X_1 \mid X_1 > \frac{H}{2})+  \Pr(X_1 \leq \frac{H}{2})\E( \max_{1\leq i \leq K} X_i \mid X_1 \leq \frac{H}{2}) \\
&\geq \frac{1-\epsilon}{2}\frac{3H}{4}+\epsilon H+\frac{1}{2}\E( \max_{2\leq i \leq K} X_i ) \\
&=\frac{3H}{8}+\frac{1}{2}\frac{K-1}{K}H+O(\epsilon).
\end{align*}
By taking $\epsilon$ to zero and applying standard manipulations, we obtain
\[
\frac{\OPT}{\OPTEA} \leq \frac{K^2}{\frac{7}{8}K^2+\frac{3}{8}K-\frac{1}{2}}.
\]
This term attains $\frac{16}{15}$ for $K=2$ and is monotonically increasing for $K\geq 3$; hence, the claim is proven by taking $K$ to infinity.
\end{proofof}

\section{Incentive Compatible Mechanism for Strategic Agents and Uniform Arrival}\label{sec:ic and uniform}
In this section, we consider strategic agents and uniform arrival. Formally, we assume that the $n$ agents arrive in a random order, $\sigma:\{1, \ldots, n\} \to \{1, \ldots, n\}$, where $\sigma$ is selected uniformly at random from the set of all permutations. We show that \mainalg satisfies IC as is, assuming that there are sufficiently many agents. We introduce the following quantity $\delta$. Let $\delta_i = \Pr(\forall i'\in [K]\setminus \{i\} : X_i > X_{i'} )$. In words, $\delta_i$ is the probability that arm $a_i$ is superior to all other arms. Clearly, if Assumption \ref{assumption ic} holds, $\delta_i >0$ for every arm $i\in [K]$. In addition, let $\delta = \min_{i\in [K]} \delta_i$. Lemma \ref{FEE is IC} implies that if there are $poly(H,K,\frac{1}{\delta})$ agents, then \mainalg is IC.
\begin{lemma}\label{FEE is IC}
Under Assumption \ref{assumption ic} and uniform arrival, if $n\geq \frac{24H^2}{\delta} \max\left\{K,H \ln {\frac{4H}{\delta}}\right\} $, then \textnormal{\mainalg} is IC.
\end{lemma}

\begin{proofof}{Lemma \ref{FEE is IC}}
To prove the statement, we need to show that whenever an agent is recommended arm $a_r$, her best response is to select arm $a_r$. We focus on an arbitrary agent, and present the analysis from her point of view. In addition, if $r=1$, either she is the first agent to arrive at the system or no better arm was discovered, resulting in $a_1$ being a best response. Otherwise, $r\neq 1$. We define the following events: let $E^r_{rec}$ be the event indicating that FEE recommends arm $a_r$ to the agent; $E^r_{open}$ indicates whether arm $a_r$ was recommended to \textit{some} agent; and $E^r_{opt}$ indicates whether $a_r$ is an optimal arm. All of these events are defined w.r.t. the distribution over histories and the agent arrival distribution. Due to the uniform arrival distribution, the probability of $E^r_{rec}$ matches the proportion of agents who are recommended arm $a_r$. We proceed by analyzing the odds of being recommended $a_r$. Due to the definition of $\epsilon$ and the way \mainalg works when it observed a superior arm,
\begin{equation}\label{eq:opt bounds}
\Pr\left(\eopt \mid \eopen\right) \geq \delta, \quad  \Pr\left(\overline \eopt \mid \eopen\right) \leq 1-\delta.
\end{equation}
Next, we present a lemma that gives a large deviation bound on the number of agents needed for the experience phase.
\begin{lemma}\label{lemma:not too many}
Let $Q(\epsilon)=\max\{2KH,{2H^2}\ln {\frac{1}{\epsilon}}\}$. The experience phase terminates after $Q(\epsilon)$ agents w.p. of at least $1-\epsilon$.
\end{lemma}
The proof of Lemma \ref{lemma:not too many} and other claims we use in this lemma appear just after the end of this proof. For simplicity, denote $Q=Q(\epsilon)$. Conditioning  on $\eopen$, arm $a_r$ is either recommended exactly once (in case its reward is observed to be inferior to another arm during the execution), or several times. The latter can only happen if $R_r>R_1$ and arm $a_r$ is used by \mainalg to explore other, unobserved arms. In this case, Lemma \ref{lemma:not too many} implies that would not happen more than $Q$ times, w.h.p. As a result,
\begin{equation}\label{eq:the agent gets r notopt}
\Pr(\eal \mid \overline \eopt, \eopen) \leq 
(1-\epsilon)\frac{Q+1}{n} +\epsilon \frac{n}{n} \leq \frac{Q+1+\epsilon n}{n}.
\end{equation}
Moreover,
\begin{observation}\label{observation ar is the best}
For every history $h$ such that $\eopt,\eopen$ occur, if \textnormal{\algname} already reached the exploit phase (Line \ref{alg: best arm}) under $h$, then $\Pr(\textnormal{\mainalg}(h)=a_r)=1$.
\end{observation}
Due to Observation \ref{observation ar is the best}, we also conclude that
\begin{equation}\label{eq:the agent gets r opt}
\Pr(\eal \mid  \eopt, \eopen) \geq 
(1-\epsilon)\frac{n-Q-K}{n}.
\end{equation}
We now analyze the ratio between the probability of arm $a_r$ being optimal and the probability that it is not, given $\eal$. We have
\begin{align}\label{eq:ratio bound}
\frac{\Pr( \eopt \mid  \eal, \eopen)}{\Pr(\overline \eopt  \mid \eal, \eopen)} &= \frac{\frac{\Pr(\eal , \eopt, \eopen)}{\Pr(\eal ,\eopen)}}{\frac{\Pr(\eal , \overline \eopt, \eopen)}{\Pr(\eal ,\eopen)}} \nonumber \\
&= \frac{\Pr(\eal , \eopt, \eopen)}{\Pr(\eal , \overline \eopt, \eopen)} \\
&= \frac{\Pr(\eopen)\Pr(\eopt\mid \eopen)\Pr(\eal\mid \eopen,\eopt)}{\Pr(\eopen)\Pr(\overline \eopt\mid \eopen)\Pr(\eal\mid  \eopen, \overline\eopt)} \nonumber .
\end{align}
Applying the bounds from Equations (\ref{eq:opt bounds}),(\ref{eq:the agent gets r notopt}) and (\ref{eq:the agent gets r opt}) to Equation (\ref{eq:ratio bound}), we get
\begin{equation}
\frac{\Pr( \eopt \mid  \eal, \eopen)}{\Pr(\overline \eopt  \mid \eal, \eopen)}\geq \frac{\delta (1-\epsilon)\frac{n-Q-K}{n}}{(1-\delta)\frac{Q+1+\epsilon n}{n}} \nonumber,
\end{equation}
and by rearranging we obtain
\begin{equation}\label{eq:opt ration bound}
\Pr( \eopt \mid  \eal, \eopen) \geq \Pr( \overline \eopt \mid  \eal, \eopen) \frac{\delta (1-\epsilon)(n-Q-K)}{(1-\delta)(Q+1+\epsilon n)}.
\end{equation}
Next, we bound the expected difference between the reward of arm $a_r$ and that of an arbitrary arm $a_i$, with $i\neq r$. We have
\begin{align}\label{eq exp diff}
\E(X_r-X_i\mid \eal) &= \E(X_r-X_i\mid \eal, \eopen) \nonumber\\
&=  \E(X_r-X_i\mid \eal, \eopen,\eopt)\Pr( \eopt \mid  \eal, \eopen) \\
&\quad \quad + \E(X_r-X_i\mid \eal, \eopen, \overline \eopt)\Pr( \overline \eopt \mid  \eal, \eopen) \nonumber \\
&\geq  1\cdot \Pr( \eopt \mid  \eal, \eopen)-H\cdot \Pr( \overline \eopt \mid  \eal, \eopen) \nonumber.
\end{align}
By plugging in the bound obtained in Equation (\ref{eq:opt ration bound}) to Equation (\ref{eq exp diff}) we get
\begin{align}\label{eq: exp diff final}
\E(X_r-X_i\mid \eal) \geq \Pr( \overline \eopt \mid  \eal, \eopen)\left( \frac{\delta (1-\epsilon)(n-Q-K)}{(1-\delta)(Q+1+\epsilon n)} -H   \right).
\end{align}
Ultimately, since
\begin{observation}\label{obs:n and n prime}
Let $\epsilon=\frac{\delta}{4H}$ and $Q=\max\{2KH,{2H^2}\ln {\frac{4H}{\delta}}\}$. If $n\geq \frac{6HQ}{\delta} $, it holds that
\[
\frac{\delta (1-\epsilon)(n-Q-K)}{(1-\delta)(Q+1+\epsilon n)}\geq H.
\]
\end{observation}

The proof is completed by combining Observation \ref{obs:n and n prime} with Equation (\ref{eq: exp diff final}) to show that $\E(X_r-X_i\mid \eal)\geq 0$ for every arm $a_i$.

\end{proofof}

\begin{proofof}{Lemma \ref{lemma:not too many}}
Let $Z$ denote the number of agents receiving recommendations in the experience phase (Lines \ref{alg:discover new arm} and \ref{alg:reuse good arm}). 
The proof is based on two observations: first, we show that $Z$ is first-order stochastically dominated by an easy-to-analyze random variable. Then, we use a concentration bound to complete the proof.
\begin{observation}\label{prop:dominated}
For every $z\in \mathbb N$,
\[
\Pr\left(\textnormal{NBin}(K, \frac{1}{H})\geq z\right)) \geq \Pr(Z \geq z).
\]
\end{observation}
Moreover, using Hoeffding's inequality we have
\begin{claim}\label{claim: negative binom}
Let $\epsilon>0$, $K,H \in \N$, and let $Q=\max\{2KH,{2H^2}\ln {\frac{1}{\epsilon}}  \}$ . It holds that
\[
\Pr\left(\textnormal{NBin}(K, \frac{1}{H}) \geq Q \right) \leq \epsilon.
\]
\end{claim}
By combining Observation \ref{prop:dominated} and Claim \ref{claim: negative binom}, we get 
\[
\Pr(Z \geq Q) \leq \Pr(\textnormal{NBin}(K, \frac{1}{H})\geq Q)\leq \epsilon
\]
This completes the proof of this lemma.
\end{proofof}

\begin{proofof}{Observation \ref{observation ar is the best}}
To see why Observation \ref{observation ar is the best} holds, recall that if $\eopen$ occurs, then \mainalg revealed $R_r$. Moreover, reaching Line \ref{alg: best arm} suggests that the experience phase is over; therefore, the rewards of all arms are revealed. Finally, since $\eopt$ holds, \mainalg will pick it with probability 1.
\end{proofof}

\begin{proofof}{Observation \ref{obs:n and n prime}}
First, notice that $\epsilon < \frac{1}{2}$ and $Q>K$; thus,
\begin{equation}\label{eq: obs simple change}
\frac{\delta (1-\epsilon)(n-Q-K)}{(1-\delta)(Q+1+\epsilon n)}\geq \frac{\frac{\delta}{2}(n-2Q)}{(2Q+\epsilon n)}.
\end{equation}
It suffices to show that the right-hand side of Equation (\ref{eq: obs simple change}) is greater or equal to $H$. Now,
\begin{align}
& \frac{\frac{\delta}{2}(n-2Q)}{(2Q+\epsilon n)} \geq H \Leftrightarrow  \frac{\delta}{2}(n-2Q) \geq H (2Q+\epsilon n) \Leftrightarrow  \frac{\delta n}{2} -\delta Q \geq 2HQ +\epsilon H n \nonumber \\
&\Leftrightarrow  \frac{\delta n}{2} - \epsilon H n  \geq 2HQ +\delta Q \Leftrightarrow  n\left( \frac{\delta}{2} - \epsilon H \right)   \geq 2HQ +\delta Q \Leftrightarrow  n   \geq \frac{Q(2H+\delta)}{\left( \frac{\delta}{2} - \epsilon H \right)} .
\end{align}
Inserting the values of $\epsilon$ and $Q$, we argue that the statement holds as long as
\begin{align}\label{eq: n is greater}
n   \geq \frac{\max\{2KH,{2H^2}\ln {\frac{4H}{\delta}}\}(2H+\delta)}{\left( \frac{\delta}{2} - \frac{\delta}{4H} H  \right)}=\frac{4\max\{2KH,{2H^2}\ln {\frac{4H}{\delta}}\} (2H+\delta)}{\delta}.
\end{align}
To conclude the proof, recall that $n\geq \frac{12HQ}{\delta}$; hence
\[
n \geq  \frac{12H\max\{2KH,{2H^2}\ln {\frac{4H}{\delta}}\}}{\delta} \geq\frac{4\max\{2KH,{2H^2}\ln {\frac{4H}{\delta}}\} (2H+\delta)}{\delta};
\]
thus, Equation (\ref{eq: n is greater}) holds.
\end{proofof}
\begin{proofof}{Claim \ref{claim: negative binom}}
First, observe that
\begin{equation}\label{eq cl step 1}
\Pr\left(\textnormal{NBin}(K, \frac{1}{H})  \geq Q \right) = \Pr\left(\textnormal{Bin}(Q, \frac{1}{H}) \leq  K \right).
\end{equation}
Next, notice that $k\leq \frac{Q}{2H}$; thus, 
\begin{equation}\label{eq cl step 2}
\textnormal{Eq.} (\ref{eq cl step 1}) \leq \Pr\left(\textnormal{Bin}(Q, \frac{1}{H}) \leq   \frac{Q}{2H} \right).
\end{equation}
By using the multiplicative version of the Chernoff Bound, we get that 
\begin{equation} \label{eq cl step 3}
\textnormal{Eq.} (\ref{eq cl step 2}) \leq e^{-\frac{Q}{2H^2}}.
\end{equation}
Recall that $Q\geq {2H^2}\ln {\frac{1}{\epsilon}}$; therefore,
\[
e^{-\frac{Q}{2H^2}} \leq e^{-\frac{{2H^2}\ln {\frac{1}{\epsilon}}}{2H^2}}=\epsilon.
\]
\end{proofof}

\begin{proofof}{Observation \ref{prop:dominated}}
The exploration phase of \mainalg is based on $\pi^*$. Once $\pi^*$ reaches a terminal state, there are two options:
\begin{itemize}
\item The terminate state exhibits $\beta=R_1$. In this case, the statement of the If sentence in Line \ref{alg:else towards experience} is false, and there is no need for experience. Consequently, $Z=0$ w.p. 1 and the statement holds.
\item The terminate state exhibits $\beta>R_1$. In this case, \mainalg  enters the While loop in Line \ref{alg:while loop}. In each iteration of the While loop, either the size of $U$ decreases by 1 (Lines \ref{alg: while explore dominated} and \ref{alg:discover new arm}), or stays the same (Line \ref{alg:reuse good arm}). The statement in Line \ref{alg:reuse good arm} will only execute if the arm $a_i$ selected in Line \ref{alg:select a_i} satisfies $\mu_i<R_1$, otherwise the If condition in Line \ref{alg:while if conditions} would execute; hence, the probability of executing Line \ref{alg:reuse good arm} is bounded by 
\[
\Pr\left(Uni(0,1)\geq  \frac{R_{\tilde i}-R_1}{R_{\tilde i}-\mu_i}\right)\leq \Pr\left(Uni(0,1)\geq  \frac{1}{H}\right)=1-\frac{1}{H}.
\]
This applies for every iteration of the While loop. Recall that there are at most $K-2$ arms needed to be explored, and hence the statement holds.
\end{itemize}
\end{proofof}

\subsection{The Full Exploration Mechanism}
\begin{proposition}\label{ic mech}
Under Assumption \ref{assumption ic} and uniform arrival, \fulexp~ is IC and asymptotically optimal.
\end{proposition}
\begin{proofof}{Proposition \ref{ic mech}}
Asymptotic optimality is straightforward. The proof of being IC goes along the lines of Theorem~\ref{FEE is IC} and hence omitted.
\end{proofof}

\section{Elaborated Example of \mainalg}\label{sec:elaborated example.}
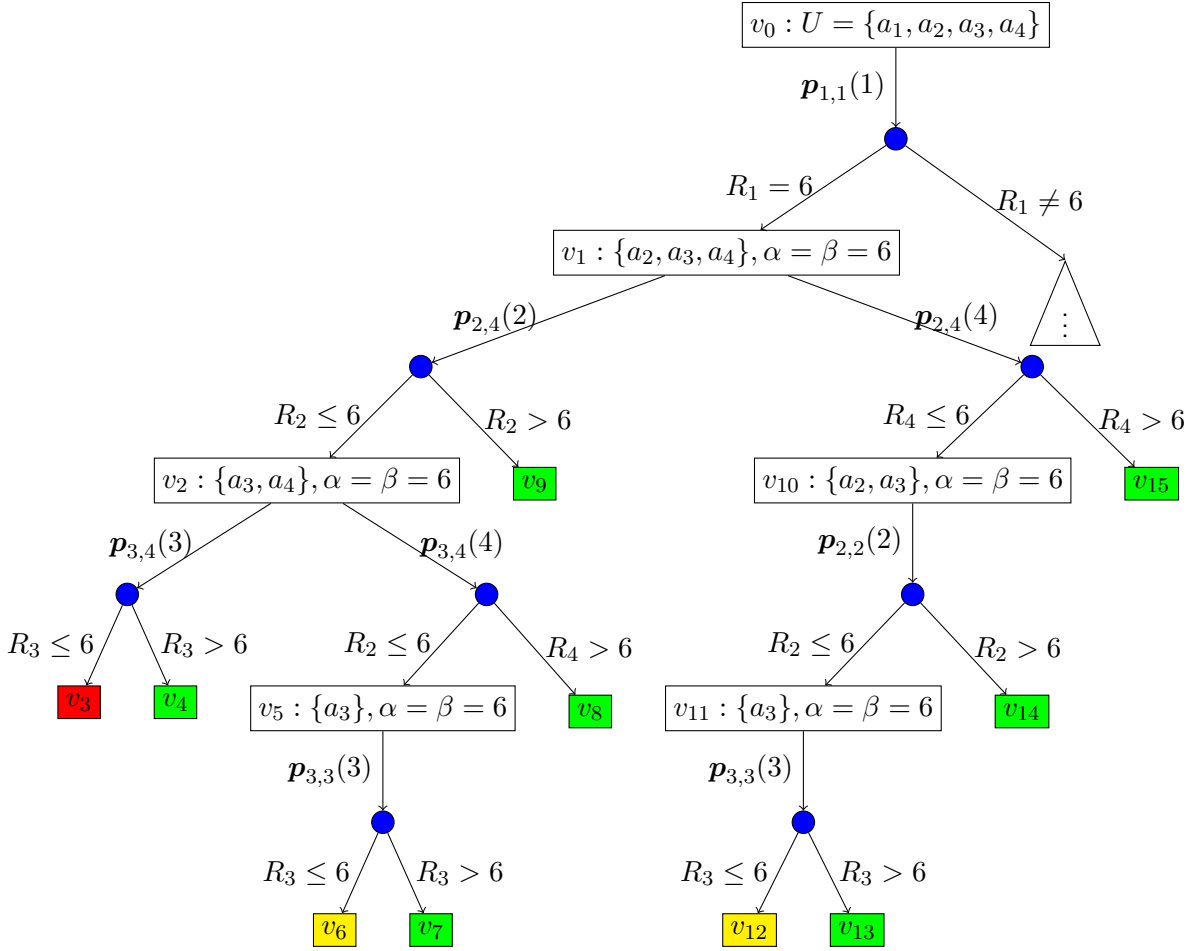
\begin{figure}[t]
\centering
\usetikzlibrary{shapes.geometric}
\forestset{
 strongedge label/.style 2 args={
    edge label={node[midway,left, #1]{#2}},
  }, 
 weakedge label/.style 2 args={
    edge label={node[midway,right, #1]{#2}},
  }, 
   noedge label/.style 2 args={
    edge label={color=white,node[midway,right, #1]{#2}},
  }, 
  important/.style={draw={red,thick,fill=red}},
  goodt/.style={draw={yellow,thick,fill=green}},
  ssarbre/.style={isosceles triangle,
                  draw,
                  shape border rotate=90,
                  minimum size=.7cm,
                  child anchor=apex,
                  anchor=apex}
}
\begin{forest} baseline,for tree=
	{
	draw,
	font=\sffamily,
	l+=.2cm,
	inner sep=3pt,
	l sep=30pt,
	s sep=20pt,
	edge={->},
    }
	[{$v_0:U=\{a_1,a_2,a_3,a_4\}$},
		[,strongedge label={}{$\bl p_{1,1}(1)$}, fill=blue, circle
		 [{$v_1:\{a_2,a_3,a_4\},\alpha=\beta=6$}, strongedge label={}{$R_1=6$}
			[,strongedge label={}{$\bl p_{2,4}(2)$}, fill=blue, circle
				[{$v_2:\{a_3,a_4\},\alpha=\beta=6$}, strongedge label={}{$R_2\leq6$},
					[, strongedge label={}{$\bl p_{3,4}(3)$}, fill=blue, circle
						[{$v_3$}, strongedge label={}{$R_3\leq 6$},fill=red
						]
						[{$v_4$}, weakedge label={}{$R_3>6$},fill=green
						]
					]
					[, weakedge label={}{$\bl p_{3,4}(4)$}, fill=blue, circle
						[{$v_5:\{a_3\},\alpha=\beta=6$}, strongedge label={}{$R_2\leq 6$}, name=node vfive
							[, strongedge label={}{$\bl p_{3,3}(3)$}, fill=blue, circle
								[{$v_6$}, strongedge label={}{$R_3\leq6$},fill=yellow
								]
								[{$v_7$}, weakedge label={}{$R_3>6$},fill=green
								]	
							]							
						]
						[{$v_8$}, weakedge label={}{$R_4>6$},fill=green
						]			
					]
				]
				[{$v_9$}, weakedge label={}{$R_2>6$},fill=green
				]
			]
			[, weakedge label={}{$\bl p_{2,4}(4)$}, fill=blue, circle
				[{$v_{10}:\{a_2,a_3\},\alpha=\beta=6$}, strongedge label={}{$R_4\leq6$},
					[, strongedge label={}{$\bl p_{2,2}(2)$}, fill=blue, circle
						[{$v_{11}:\{a_3\},\alpha=\beta=6$}, strongedge label={}{$R_2\leq 6$}
							[, strongedge label={}{$\bl p_{3,3}(3)$}, fill=blue, circle
								[{$v_{12}$}, strongedge label={}{$R_3\leq6$},fill=yellow
								]
								[{$v_{13}$}, weakedge label={}{$R_3>6$},fill=green
								]	
							]			
						]
						[{$v_{14}$}, weakedge label={}{$R_2>6$},fill=green
						]
					]
				]	
				[{$v_{15}$}, weakedge label={}{$R_4>6$},fill=green
				]
			]
		 ]
		 [\vdots,ssarbre, weakedge label={}{$R_1 \neq 6$}]
		]
	]
\end{forest}
\caption{
Visualization of $\pi^*$ obtained for the example in Section \ref{sec:elaborated example.}. The right child of $v_0$ encapsulates the sub-tree of the policy for $R_1 \neq 6$.
 \label{fig:example-tree}}
\end{figure}

In this section, we provide an elaborated example of the way \mainalg operates. Consider $K=4$ arms, $H=40$ and $X_1\sim Uni\{0,\dots 40\}, X_2\sim Uni\{0,\dots 30\}, X_3\sim Uni\{0,\dots 20\}, X_4\sim Uni\{0,\dots 10\}$; 
thus $\mu_1=20$, $\mu_2=15$, $\mu_3=10$, and $ \mu_4=5$. As always, $a_1$ is the default arm. Let us assume for the sake of this example that $\bl X = (X_1,X_2,X_3,X_4)=(6,3,7,2)$, but these values are not known to the algorithm. We illustrate $\pi^*$ in Figure \ref{fig:example-tree}, obtained from a simple Python program.

Nodes with a square frame are associated with states of the GMDP. The leaves are terminal states, and the intermediate nodes are non-terminal. Blue circled nodes are auxiliary, and separate between values the newly observed arm can take. The outgoing edges from each non-terminal white node are the transition probabilities. For instance, in $v_1$, the outgoing edges are $\bl p_{2,4}(2)$ and $\bl p_{2,4}(4)$, hinting that the action taken in $v_1$ is $\bl p_{2,4}$. 

The colored leaves represent terminal states. Green leaves are terminal states where the policy revealed an arm with a value greater than $R_1$, i.e., $\beta >R_1 $ (see Line \ref{alg:else towards experience} in \mainalg). Yellow leaves are terminal states in which $\pi^*$ reveals all the rewards, but those are less or equal to $R_1$. And the red node, $v_3$ refers to the terminal state in which $a_2,a_3$ were explored and $R_2,R_3$ were less or equal to $R_1$, and $a_4$ was not explored. Notice that $v_5$ and $v_{11}$ are associated with the same state, and since $\pi^*$ is stationary, their sub-trees are identical. Per our assumption on $\bl X$, the GMDP will reach one of the leaves in $\{v_4,v_7,v_{13}\}$, depending on the coin flips. To illustrate, we assume that $\pi^*$ reached $v_4$ and explain the trajectory.

The root of the tree, $v_0$, denotes the initial state $s_0$. Due to the construction of the optimal policy $\pi^*$, it will always explore $a_1$ in the first round (level 0 of the tree in Figure \ref{fig:example-tree}); thus, \mainalg recommends the first agent $a_1$, and observes that $R_1=6$ (recall we assume the rewards are according to $\bl X$). The GMDP then transitions to $v_1$. At $v_1$, $\pi^*$ picks $\bl p_{2,4}$. \mainalg then draw coins (Line \ref{alg:using policy}), which realized with $a_2$ (since we assume the leaf $v_4$ was realized eventually), and selects $a_2$ for the second agent. The value of $R_2=3$ is then observed, and the GMDP moves $v_2$. $\pi^*$ picks $\bl p_{3,4}$, \mainalg draw coins (Line \ref{alg:using policy}), which realized with $a_3$, and selects $a_2$ for the second agent. The value of $R_3=7$ is realized, and the GMDP reaches $v_4$, which is a terminal state. \mainalg exists the while loop in Line \ref{alg:while not terminal}. \mainalg then enters the if statement of Line \ref{alg:else towards experience}, since it observe that $\beta=R_3>R_1$. At this point, the set of unobserved arms $U$ is $\{a_4\}$, and so \mainalg enters the while loop in Line \ref{alg:while loop}. In Line \ref{alg: pick a r tilde}, it sets $a_{\tilde r}=a_3$, following by setting $a_i=a_4$ in the subsequent line. Since there is a positive probability that $X_4 > R_3$, \mainalg skips the if block in Line \ref{alg: dominated}.

Then, in Line \ref{alg:draw}, \mainalg draws $Y\sim Uni[0,1]$. Since $\mu_4 = 5 \leq 6=  R_1$, the second condition of the if block in Line \ref{alg:while if conditions} does not hold; hence, the only way to enter the if block in Line \ref{alg:while if conditions} is by having $Y \leq \frac{R_3-R_1}{R_3-\mu_4}=\frac{7-6}{7-5}=0.5$. If $Y >0.5$, \mainalg moves to Line \ref{alg:reuse good arm}, recommends $a_3$ to the fourth agent, and starts another iteration of the while loop in Line \ref{alg:while loop}. With probability 1, after finitely many agents, \mainalg will draw $Y \leq 0.5$. Then, it will recommend $a_4$ in Line \ref{alg:discover new arm}, and observe $R_4$. In Line \ref{alg: while explore open}, $U$ becomes the empty set. \mainalg will then exit the while loop in Line \ref{alg:while loop}, move to Line \ref{alg: best arm}, and every subsequent agent will exploit---\mainalg will recommend $a_3$ from then on.

}\fi}

\end{document}